\documentclass[%
 reprint,
nofootinbib,
 amsmath,amssymb,
 aps,
]{revtex4-1}

\usepackage{graphicx}% Include figure files
\usepackage{dcolumn}% Align table columns on decimal point
\usepackage{bm}% bold math
\usepackage{hyperref}% add hypertext capabilities
\usepackage{aas_macros}
\usepackage[capitalize]{cleveref}
\usepackage[dvipsnames,table]{xcolor}
\hypersetup{urlcolor=RedViolet,
	    citecolor=ForestGreen ,
	    linkcolor=RoyalBlue, 
	    colorlinks=true}
\usepackage[normalem]{ulem}

\newcommand{\Mpch}{$h^{-1}\,\mbox{Mpc}$}
\newcommand{\kpch}{$h^{-1}\,\mbox{kpc}$\,}

\newcommand{\ie}{\textit{i.e.}}
\newcommand{\eg}{\textit{e.g.}}

\newcommand{\smt}{SMT-01}
\newcommand{\elephant}{\textsc{elephant}}
\newcommand{\dudeg}{n\textsc{DGP-HR-1400}}
\newcommand{\mgmil}{n\textsc{DGP-HR-1280}}
\newcommand{\lcdm}{$\mathrm{\Lambda \text{CDM}}$}
\newcommand{\fsig}{F(\sigma)}
\def\dd{\textrm{d}}

\begin{document}

%\preprint{APS/123-QED}

\title{Universality of the halo mass function in modified gravity cosmologies}
\author{Suhani Gupta}
\email{gupta@cft.edu.pl}
\author{Wojciech A. Hellwing}
\author{Maciej Bilicki}
\author{Jorge Enrique Garc\'ia-Farieta}
 \affiliation{Center for Theoretical Physics, Polish Academy of Sciences, Al. Lotnik\'ow 32/46, 02-668 Warsaw, Poland }

\date{\today}% It is always \today, today,

\begin{abstract}
We study the halo mass function (HMF) in modified gravity (MG) models using a set of large $N$-body simulations -- the \elephant{} suite. We consider two popular beyond-general relativity scenarios: the Hu-Sawicki chameleon $f(R)$ model and the normal branch of the Dvali-Gabadadze-Porrati (nDGP) braneworld. We show that in MG, analytic formulation based on the Press-Schechter framework offers a grossly inaccurate description of the HMF. We find, however, that once the HMF is expressed in terms of the dimensionless multiplicity function, it approximately assumes a redshift-independent universal character for all the models.
Exploiting this property, we propose universal fits for the MG HMF in terms of their fractional departures from the \lcdm{} case. We find two enclosed formulas, one for $f(R)$ and another for  
nDGP, that provide a reliable description of the HMF over the mass range covered by the simulations. These are accurate to a few percent with respect to the $N$-body data. We test the extrapolation potential of our fits against separate simulations with a different cosmological background and mass resolution, and find very good accuracy, within $\sim 10\%$. A particularly interesting finding from our analysis is a Gaussian-like shape of the HMF deviation that seems to appear universally across the whole $f(R)$ family, peaking at a mass variance scale characteristic for each $f(R)$ variant. We attribute this behavior to the specific physics of the environmentally-dependent chameleon screening models.
\end{abstract}

\maketitle

%\tableofcontents

\section{Introduction}
\label{sec:intro}

Our current standard model of cosmology -- \textit{Lambda Cold Dark Matter} (\lcdm{}) -- is a very successful description of the evolution of the Universe from the hot relativistic Big Bang until the present time, some 13.8 billion years later. This simple 6-parameter model can account very well for the primordial nucleosynthesis light elements abundance, the precisely observed properties of the cosmic microwave background, the large-scale clustering of matter, as well as for the late-time accelerated expansion history \citep{bbn_review,Planck2018,sdss4_highz,sdss2_snls_de}.

However, \lcdm{} is of inherent phenomenological nature, due to the necessity to include dark matter and dark energy fluids in the cosmic matter-energy 
budget \citep{Planck2018,2019_des}, when general relativity (GR) is assumed as the underlying theory of gravity. The phenomenological character of \lcdm{}, together with known theoretical issues related to reconciling the absurdly small value of the cosmological constant with quantum vacuum theory predictions \citep{cc_problem}, and the observed anomalies associated with \lcdm{} \citep{lcdm_small_scale_challenges,H0_tension,lcdm_tensions_sn_quasar_grb}, have motivated searches for alternative scenarios or extensions to the concordance model. One particularly vibrant research theme in the last decade focused on attributing the accelerated late-time expansion to some beyond-GR extensions (usually scalar-tensor theories), rather than to the vanishingly small cosmological constant. Such models are commonly dubbed as ``modified gravity'' and have been put forward and studied in their many rich flavors in numerous works 
\citep[\eg{}][]{mg_cosmo,gravity_allscales_baker,cosmological_tests_mg,Nojiri2017_MGreview,cosmological_tests_ferreira,fR_gravity_theories,Tessa_Novel_Probes,beyond_gr_review,de_vs_mg,beyond_lcdm}.

It is important to note that these so-called modified gravity (MG) theories cannot be so far considered as fully-flagged competitors to Einstein's GR since they are not new independent metric theories of gravity. These are rather a collection of useful phenomenological models, exploring the freedom of modifying the Einstein-Hilbert action to produce a physical mechanism effectively mimicking the action of the cosmological constant \citep{extended_theories_gravity, mg_cosmo, de_vs_mg}. This is usually achieved by introducing some extra degrees of freedom in the space-time Lagrangian. The MG models provide a very useful framework that allows to both test GR on cosmological scales and to explore the physics of cosmic scalar fields. Most of the popular MG scenarios are constructed in such a way that they have negligible consequences at early times and share the same expansion history as \lcdm{}. Owing to this, the extra physics of MG models does not spoil the great observational success of the standard model. Most of the viable beyond-GR theories assume the same cosmological background as \lcdm{}, which sets the stage for the formation and evolution of the large-scale structure (LSS).

To satisfy the observational constraints on gravity \citep[\eg][]{gr_solarsystem,gr_pulsar,gr_em_gw,gr_gw,s2_smbh,gr_deflection_light}, namely to recover GR in high-density regimes where it is well tested, and to match  the \lcdm\ expansion history, MG theories need to be supplemented with screening mechanisms. Among these, Chameleon \citep{Khoury2003PRD} and Vainshtein \citep{Vainshtein_1972} effects are most frequently studied. The physics of these screening mechanisms is the strongest factor that differentiates various MG models, and their interplay and back-reaction with the LSS and cosmological environment can have a pivotal role in administering the MG-induced effects. Thus, a meaningful classification of various MG theories can be made based on the virtues of the screening mechanism they invoke \citep{screening_review,cosmological_tests_mg,Brax_screening_mg,astrophysical_tests_screening,beyond_gr_review,dynamical_mass_mg}.

Due to the shared cosmological background, it is in the properties of LSS, where we can expect the predictions of MG models to differ from \lcdm. The formation and evolution of the LSS is governed by gravitational instability, responsible for the aggregation of dark matter (DM) and gas from primordial fluctuations into bound clumps: DM halos, which are sites of galaxy formation. This mechanism applied on the initial Gaussian adiabatic fluctuations, described by a collisionless cold dark matter (CDM) power spectrum, yields one of the most important predictions for the structure formation: a hierarchical build-up of collapsed halos. The most fundamental characteristic of this theory is the halo mass function (HMF), which describes the comoving number density of halos of a given mass over cosmic time. Within the \lcdm{} paradigm, the HMF assumes a universal power-law shape in the low halo mass  regime, $\dd n/\dd M\sim M^{-\alpha}$, with the slope approximating $\alpha\sim 1$. This is supplemented by an exponential cut-off at the cluster and higher mass scales, and the amplitude, shape, and scale of HMF evolves with redshift \cite{PS74_MF}. However, when the HMF is expressed in %natural 
dimensionless units of the cosmic density field variance, it
assumes a universal, time-independent character
\cite[\eg][]{MF_WHITE,J01_MF,S01_MF,W06_MF,LUKIC_2007,W13_MF,D16_MF}, but see
\citep{T08_MF,C10_MF,C11_MF,B11_MF,MALLEA2021_MF} for differing results.

The HMF forms the backbone of many theoretical predictions related to late-time LSS and galaxy formation models and is widely invoked in numerous cosmological studies \citep[\eg][]{halomerger_1993, Lacey_Cole_1993, Sheth_bias, Cooray_Sheth_2002,GC_review}. Given the central role of this cosmological statistic in the study and analysis of the LSS, it is of paramount importance to revisit its universal properties and characterize its deviations from \lcdm{} in viable MG theories in both linear and non-linear density regimes. This is the main topic of our work here. 

Given the rich phenomenology of potentially viable MG theories, it would be unfeasible to explore deeply each model's allowed parameter space, whether analytically or via computer simulations. Thus, we will limit our studies to two popular cases of such MG theories. They can be both regarded as good representatives of a wider class of models exhibiting similar beyond-GR physics. The first one is the $f(R)$ gravity \citep{fR_gravity_theories}, which considers non-linear functions of the Ricci scalar, $R$, due to additional scalar fields and their interaction with matter. The second one is the normal branch of the Dvali-Gabadadze-Porrati (nDGP) model \citep{ndgp_2000}, which considers the possibility that gravity propagates in extra dimensions, unlike other standard forces. Both of these nontrivial MG theories exhibit the universal feature of a fifth force arising on cosmological scales, a consequence of extra degrees of freedom. Their gradient, expressed usually as fluctuations of a cosmological scalar field, induces extra gravitational forces between matter particles. Modifications to GR on large scales result in a different evolution of perturbations than in \lcdm{} in both linear and non-linear regime. The fifth force is expected to leave an imprint on structure formation scenarios, which would lead to testable differences in the properties of the LSS in such beyond-GR theories as compared to \lcdm{}
\citep[\eg][]{novel_probes_dm_de,cosmological_tests_gravity_jain,fR_constraint_abundance,hmf_fR_cluster,sz_fR,sz_cluster_fR,hmf_ndgp_cluster,cm_mg_1,fR_psz,hmf_fR_cluster,hmf_ndgp_cluster,Schmidt_2009_fR,Schmidt_2009_ndgp,Schmidt_Oyaizu_2008_3,fR_nonlinear_structure,ndgp_void_1,ndgp_void_2}.

We will be most interested in the non-linear and mildly non-linear regimes of such theories, where new physics can have a potentially significant impact on the formation and evolution of DM halos. To take full advantage of the wealth of data from current and upcoming surveys (DES \citep{DES_2005}, DESI \citep{DESI_2013}, EUCLID \citep{euclid_2011}, LSST \citep{LSST_2019}, to name a few), which aim to constrain the cosmological parameters, and the underlying theory of gravity to percent precision, significant efforts are required on the side of theoretical and numerical modeling to reach similar level of precision in constraining possible deviations from the \lcdm{} scenario.

As mentioned above, the HMF when expressed in scaled units is expected to exhibit a nearly universal behavior as a function of redshift. This property has been exploited extensively to devise empirical fits for the HMF, which can be then readily used for forecasting various LSS properties \citep{J01_MF, S01_MF, W13_MF, D16_MF, W06_MF}. The availability of such HMF models or simulation-based fits, precise to a few percent or better, is essential to obtain the accuracy needed for the ongoing and future LSS surveys. However, as we will elaborate below, these kinds of HMF prescriptions developed for \lcdm{} neither capture the non-linear and scale-dependent dynamics nor the screening mechanisms associated with the MG models \citep[\eg][]{Lombriser_2013, Schmidt_2009_ndgp, Schmidt_Oyaizu_2008_3}.

The inadequacy of the standard approach to HMF in MG scenarios has led to various other methodologies being developed. Among them are those based on the spherical collapse model and excursion set theory, studied in \citep{est_fR,extended_est_fR,sc_hmf_mg, Lombriser_2013, fR_nonlinear_structure,joint_hmf_mg} to formulate the HMF for $f(R)$ gravity models with chameleon screening. In Ref. \citep{fR_nonlinear_structure}, this was further extended to formulate the conditional mass function and linear halo bias, while Ref. \citep{joint_hmf_mg} included also massive neutrinos. In Ref. \cite{sphericalcollapse_braneworld}, the spherical collapse theory was used to develop the HMF and halo model in braneworld DGP scenarios. Other approaches, developed more recently, include machine-learning-based emulation techniques for modeling the non-linear regime in MG \citep{emulator_FORGE,emulator_pk_fR, emulator_hmf_aemulus,emulator_hmf_mira,emulator_pk_beyond_lcdm}.

Here we adopt a different approach to characterize the MG HMF. Instead of fitting for the absolute HMF values across redshifts and masses, we calibrate the deviation with respect to the GR trend, obtained by comparing the MG HMF with the \lcdm{} one. We have found this deviation to be universal across redshifts when expressed as a function of the cosmic density field variance ln($\sigma^{-1}$). A related approach was used in Ref. \citep{fR_constraint_abundance,fR_psz}, where the $f(R)$ HMF was computed by taking a product of the \lcdm{} HMF and a pre-factor given by the ratio of the HMF of $f(R)$ to \lcdm{}. However, these works were confined to using theoretical HMF predictions, whereas in our study we rely on the results from inherently non-linear $N$-body simulations. In this context, in Ref. \citep{cluster_abundance_chameleon_fR,hmf_ndgp_cluster} the authors also used simulation results to devise an empirical fit for the $f(R)$ and nDGP deviation w.r.t. \lcdm{}, but all these above mentioned works do not exploit the universality trend in the deviation of MG, which we address here. 

The paper is organized as follows. In section \ref{sec:mg_mod_sims}, we describe the simulations and MG models under consideration. In section \ref{sec:modeling_mf}, we elaborate on the HMF both from simulations and analytical fitting functions. In section \ref{sec:universality_tests}, we explore the mass function universality in both \lcdm{} and MG models, while section \ref{sec:fsigma_mg} is devoted to the method we devised to find MG HMF. In section \ref{sec:test_fits}, we extend our work to other simulation runs to check the reliability of our approach. Final section \ref{sec:conclusions_discussions} includes our conclusions, discussion, and future work prospects. In the Appendix, we discuss the additional re-scaling of the scales needed for the case of nDGP gravity models in our analysis.

\section{Modified gravity models and $N$-body simulations}
\label{sec:mg_mod_sims}
Our analysis focuses on dark matter halo catalog data generated using the \elephant{} (Extended LEnsing PHysics using ANalaytic ray Tracing \cite{ndgp_void_1, ALAM2020_ELEPHANT}) cosmological simulation suite. This $N$-body simulation series was designed to provide a good test-bed for models implementing two most frequently studied screening mechanisms: Chameleon \cite{Khoury2003PRD} and Vainshtein \cite{Vainshtein_1972} effects. These two ways of suppressing the fifth force are both extremely non-linear and have fundamental physical differences. The Chameleon mechanism makes a prospective cosmological scalar field significantly massive in high-density regions by inducing an effective Yukawa-like screening, and the effectiveness of the Chameleon depends on the local density; thus it induces environmental effects in the enhanced dynamics. The Vainshtein screening mechanism, on the other hand, makes the scalar field kinetic terms very large in the vicinity of massive bodies and as a result, the scalar field decouples from matter and the fifth force is screened. Vainshtein screening depends only on the mass and distance from a body and shows no explicit dependence on the cosmic environment. An elaborate and exhaustive description of these screening mechanisms is discussed in e.g.   \citep{screening_review,Khoury2003PRD,HS_fR_2007,Vainshtein_1972,nonlinear_interactions_nDGP,Vainshtein_GW170817,Brax_screening_mg,RSD_BIAS,astrophysical_tests_screening,cosmological_tests_mg}.

In our work, we consider the following cosmological models: \lcdm{}, the Hu-Sawicki $f(R)$ model with Chameleon screening   \cite{HS_fR_2007}, and the normal branch of the Dvali-Gabadadze-Porrati (nDGP) model    \cite{ndgp_2000} with Vainshtein screening. The parameter space of these MG models is sampled to vary from mild to strong linear-theory level differences from the \lcdm{} case. We consider three $f(R)$ variants with its free parameter $|f_{R0}|$ taken to be $10^{-6},10^{-5}$, and $10^{-4}$ (increasing order of deviation from \lcdm) dubbed as $f6$, $f5$ and $f4$, respectively, and two variants of the nDGP model, with the model parameter $r_c H_0=5$ and $1$ (again in increasing order of departure from \lcdm), marked consequently as nDGP(5) and nDGP(1), respectively.

The simulations were run from $z_\text{ini}= 49$ to $z_\text{fin} = 0$ employing the \textsc{ecosmog} code \citep{ECOSMOG_1,ECOSMOG_2,ECOSMOG_V_1,simulations_vainshtein}, each using $1024^3$ $N$-body particles in a 1024 \Mpch\ box. The mass of a single particle and the comoving force resolution were $m_p= 7.798 \times 10^{10} M_{\odot}h^{-1}$ and $\varepsilon=15$\kpch, respectively. Each set of simulations has five independent realizations, except for $f4$\footnote{For $f4$ we have two realizations at $z = 0$, two at $z = 0.3$, four at $z = 0.5$ and three at $z = 1$}. All these simulations were evolved from the same set of initial conditions generated using the Zel'dovich approximation \cite{Zeldovich_1970}.  A high value of the initial redshift ensures long enough time for the evolution of the system to wipe out any transients that would affect the initial particle distribution, which is a consequence of employing the first-order Lagrangian perturbation theory \cite{LPT_TRANSIENT_SCOCCIMARO}. The cosmological parameters of the fiducial 
background model were consistent with the WMAP9 cosmology     \cite{WMAP9}, namely $\Omega_{m}= 0.281$ (fractional matter density), $\Omega_{b} = 0.046$ (fractional baryonic density), $\Omega_{\Lambda} = 0.719$ (fractional cosmological constant density), $\Omega_{\nu} = 0$ (relativistic species density), $h = 0.697$ (dimensionless Hubble constant), $n_{s} = 0.971$ (primordial spectral index), and $\sigma_{8} = 0.820$ (power spectrum normalization).  These parameters apply to background cosmologies in both MG and \lcdm{} simulations.

For further processing, we take simulation snapshots saved at $z = 0, 0.3, 0.5$ and $1$. For each epoch, we analyzed the $N$-body particle distribution with the \textsc{ROCKSTAR} halo finder \cite{ROCKSTAR} to construct dark matter halo catalogs. Halos are truncated at the $R_{200c}$ boundary, which is a distance at which the enclosed sphere contains an overdensity equal to 200 times the critical density, $\rho_{crit}\equiv 3H_{0}^{2}/8\pi G$, and the corresponding enclosed halo mass is $M_{200c}$. We restrict our analysis to halos with at least 100 particles to avoid any shot noise or resolution effects, which sets the minimum halo mass in our catalog to $M_{min}= 8.20 \times 10^{12} M_{\odot}h^{-1}$. Therefore, the halo mass range we study typically covers galaxy groups and clusters.

For additional tests we have worked with two extra nDGP(1) runs based on Planck15 cosmology   \citep{Planck15}: one realization of \mgmil{} in a simulation box of size $100$ \Mpch, with $1280^{3}$ particles of mass $m_{p} = 4.177 \times 10^{7} M_{\odot}h^{-1}$ and one realization of \dudeg{} in a box of size $1000$ \Mpch, with $1400^{3}$ particles of mass $3.192 \times 10^{10} M_{\odot}h^{-1}$.

%%%%%%%%%%%%%%%%%%%%%%%%%%%%%%%%%%%%%%%%%%%%%%%%%%%%%%

\section{Halo mass function}
\label{sec:modeling_mf}

\begin{figure*}
\includegraphics[width=0.4\textwidth]{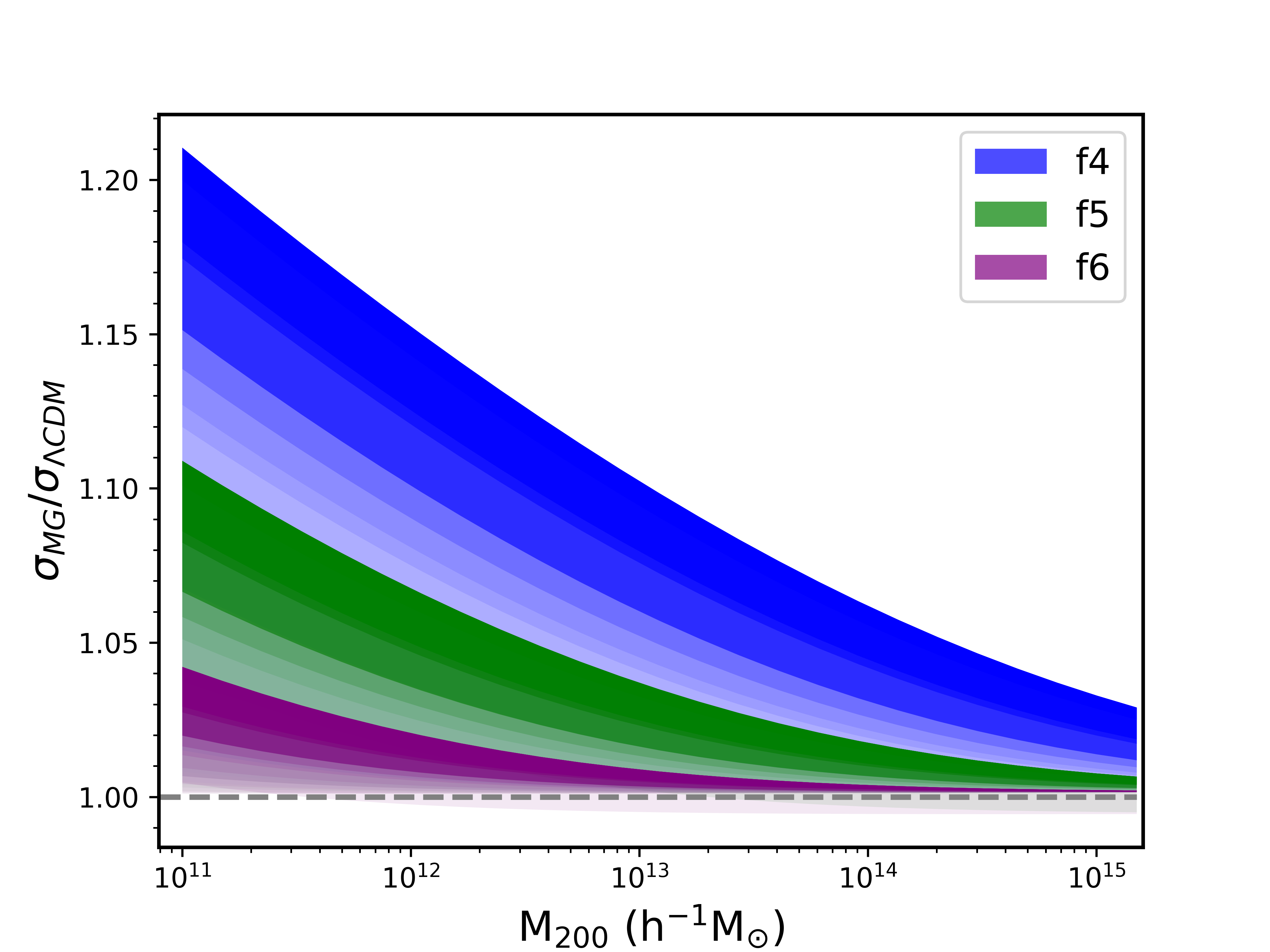}
\includegraphics[width=0.4\textwidth]{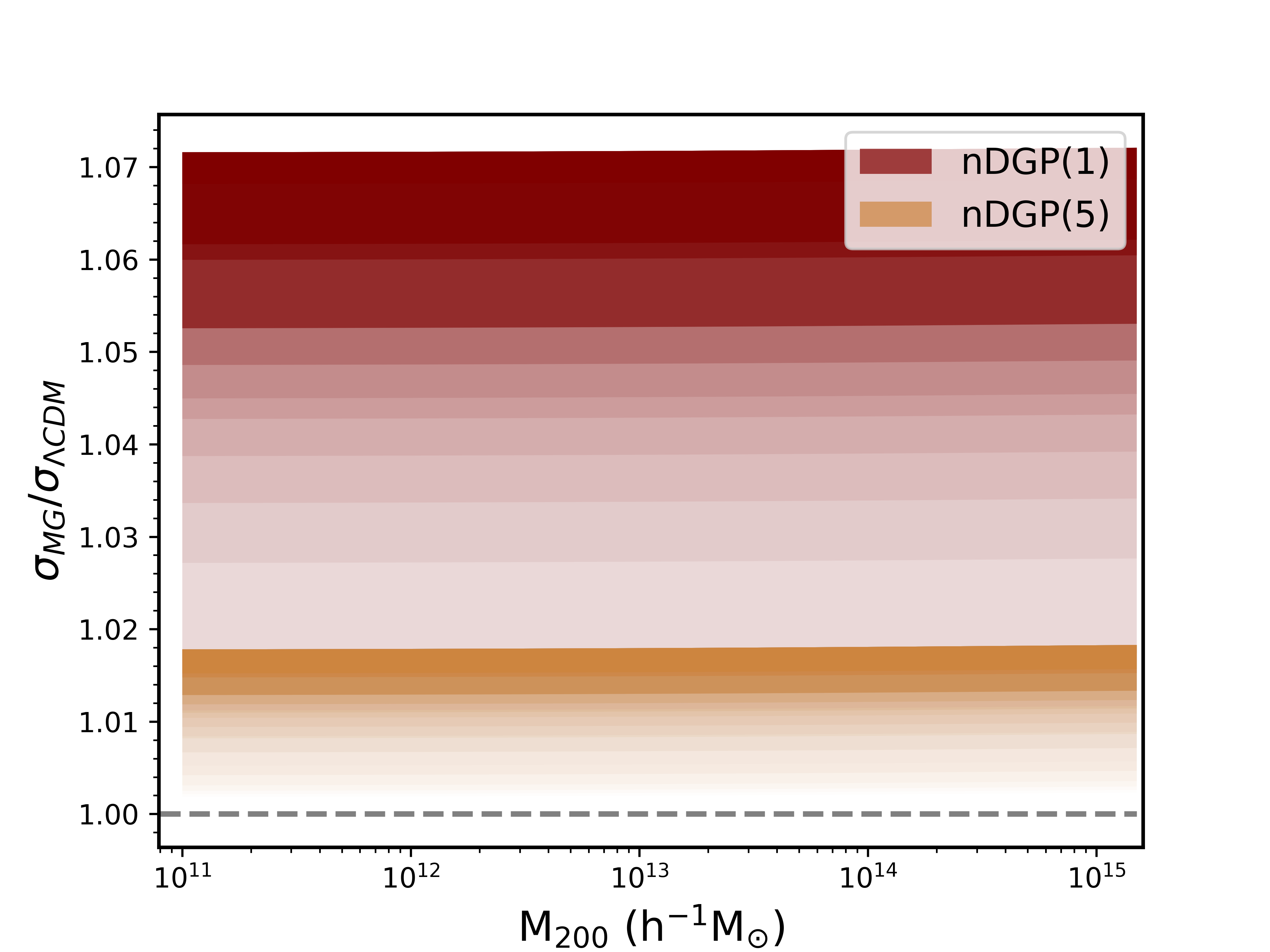}
\includegraphics[width=0.15\textwidth]{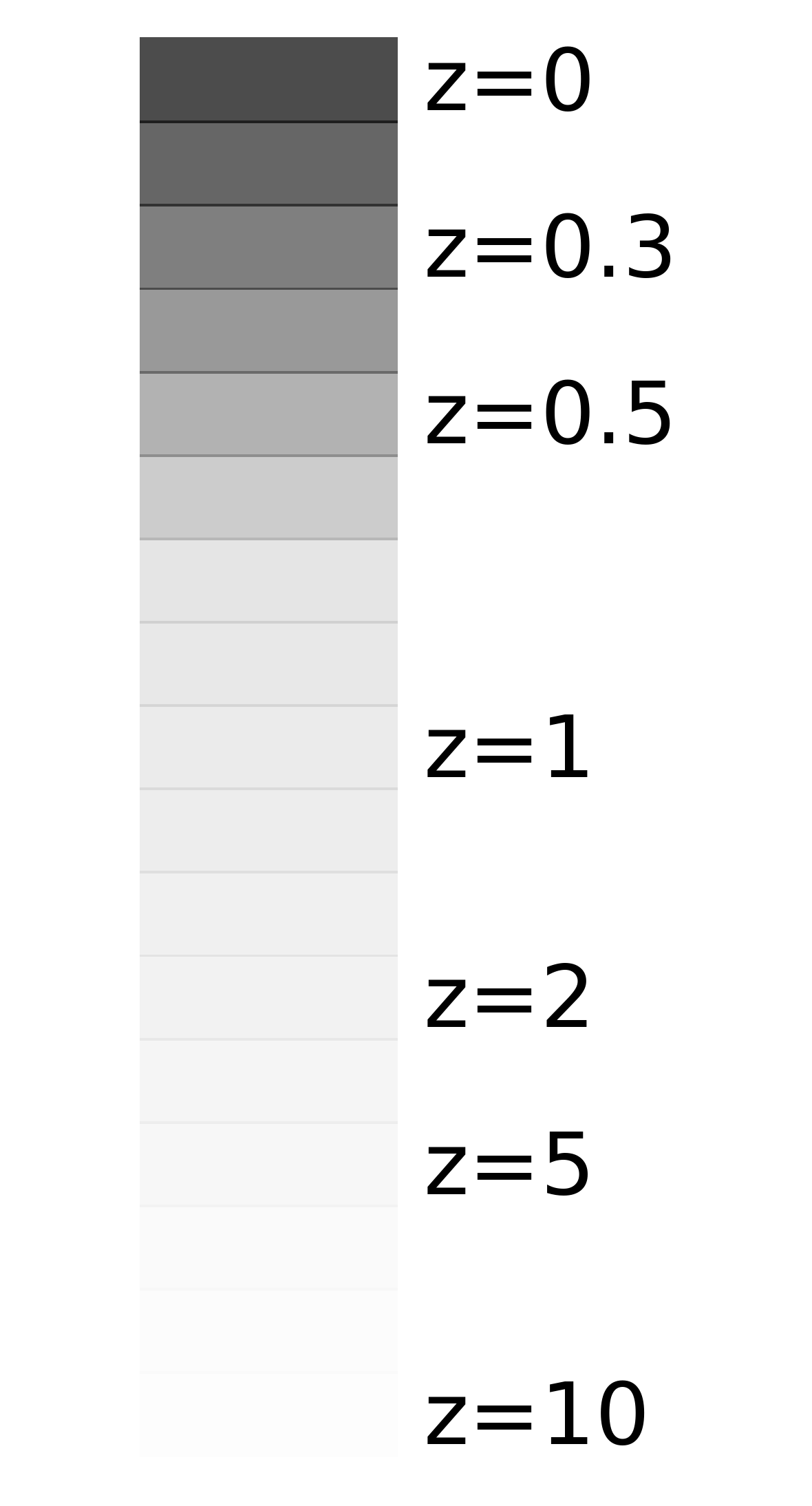}
 \caption{Ratio of the MG to \lcdm{} linear theory density variance $\sigma(M)$, defined in \cref{eqn:SIGMA_EQN}, for three variants of $f(R)$ (\textit{left-hand plot}), and two variants of the nDGP gravity model (\textit{right-hand plot}). The color gradient ranges from the darkest at $z=0$ to the lightest at $z=10$.}
  \label{fig:sigma_allz}
 \end{figure*}

The halo mass function (HMF), $n(M)$, quantifies the comoving number density of dark matter halos: their abundance is expressed as a function of halo (virial) mass at a given epoch (redshift) normalized to a unit volume. We can make a further distinction: the differential mass function $\text{d}n(M)/\text{d}M$, and its cumulative variant: $n(<M)$. In this work, we have considered the former for our analyses.

The amplitude of the HMF depends both on the matter power spectrum at a given redshift, $P(k,z)$, as well as on the background cosmology. The analytical modeling of this statistical quantity has a long history, which dates back to the seminal paper of Press \& Schechter \cite{PS74_MF} in which they formulated a simple theory of a fixed barrier in the Gaussian density fluctuation field. The collapse of structures of different sizes is modeled by applying density smoothing using a spherical function of a given comoving scale $R$. The collapse of a peak of size $R$ with mass $M$ occurs when the enclosed overdensity surpasses the critical density threshold\footnote{This is the standard spherical collapse threshold value obtained in GR   \cite{peebles_1980}.}, $\delta_{c} \simeq 1.686$. In essence, the Press-Schechter (PS) theory is an application of the spherical collapse    \cite{spherical_collapse_1972} to the cosmological Gaussian random density fluctuations. 

In this approach, halos are considered as statistical fluctuations in the Gaussian random field. For a spherical halo (peak) of mass $M$ and background matter density $\rho_{m}(z)$, the linear variance in the density fluctuation field smoothed using a top-hat filter is:
\begin{equation}
\label{eqn:SIGMA_EQN}
    \sigma^{2}(R_L,z) = \frac{1}{2\pi^{2}} \int_{0}^{\infty} k^{2} W^{2}(kR_L)  P(k, z) dk.
\end{equation}
Here, $P(k, z)$ is the linear theory matter power spectrum at a given redshift, $W(x)= 3(\sin x-x \cos x)/x^{3}$ is the Fourier counterpart of the top-hat window function, and $R_L$ is the halo Lagrangian radius, which is the smoothing radius of the filter scale, given by:
\begin{equation}
    \label{eqn:lagrangian_radius}
    R_{L} = \left(\frac{3M}{4\pi\rho_{m}}\right)^{1/3}.
\end{equation}

Now, the differential HMF can be expressed as \cite{PS74_MF,J01_MF,HMF_Peacock1990}:
\begin{equation}
\label{eqn:HMF_MAIN}
\frac{dn}{dM} = \frac{\rho_{m}}{M^2} F(\sigma) \left|\frac{d\ln \sigma}{d\ln M} \right|.
\end{equation}
From the above equations, we can see that the HMF can be related to fluctuations in the matter density field by employing the so-called \textit{halo multiplicity function}, $\fsig$    \cite{J01_MF}, which describes the mass fraction in the collapsed volume. The original PS model was amended later by the excursion set approach (also termed as the extended Press-Schechter formalism, see \eg \cite{Bond_1991}), and the following functional form of $\fsig$ was postulated:
\begin{equation}
    \label{eqn:PS_f_sigma}
    F_\text{PS}(\sigma)=\sqrt{\frac2\pi}{\frac{\delta_c}{\sigma}}\exp\left(-\frac{\delta_c^2}{2\sigma^2}\right)\,.
\end{equation}
For such a simple model, the PS HMF showed remarkably good consistency with the simulation results, especially in the intermediate halo mass regime. The onset of  precision cosmology, accompanied by the rapid growth of both size and resolution of $N$-body simulations allowed for a robust numerical estimation of HMF across many orders of magnitude    \cite{T08_MF,coco_sim_WAH,C10_MF,C11_MF,W13_MF,heitmann_2019,millennium2,A12_MF,IllustrisTNG}. These have indicated that the original $F_\text{PS}(\sigma)$ grossly over-predicts the abundance of low-mass halos, simultaneously underestimating the number of the very massive ones, in the cluster mass regime. A number of amended models have been put forward, and the literature of this subject is very rich \cite[see \eg][]{S01_MF,W06_MF,R07_MF,T08_MF,C10_MF,HMF_Peacock1990,C11_MF,W13_MF,D16_MF}. All of these alternatives have their pros and cons and usually vary with performance across masses, redshifts, and fitted cosmologies  \cite{HMFCALC_Murray_2013}. 

 We have tested many different HMF models, and found that the majority of them have very similar accuracy. For brevity, we take one particular model as our main choice for the analytical HMF predictions. We use the $\fsig$ formula proposed by Sheth, Mo \& Tormen     \cite{S01_MF} (hereafter \smt{}), given as:
\begin{equation}
    F_\text{\smt}(\sigma)=A\sqrt{\frac{2a}{\pi}}\left[1+\left(\frac{\sigma^2}{\delta_c^2 a}\right)^{p}\right]{\frac{\delta_c}{ \sigma}}\exp\left(-\frac{a\delta_c^2}{2\sigma^2}\right).
\end{equation}
Here, the constants $A=0.3222$, $a=0.707$, and $p=0.3$ were found by relaxing the PS assumptions %of spherical symmetry of a collapsing peak and a fixed barrier,
and allowing for an ellipsoidal peak shape along with the possibility of a moving barrier.

\begin{figure*}
\includegraphics[width=\linewidth]{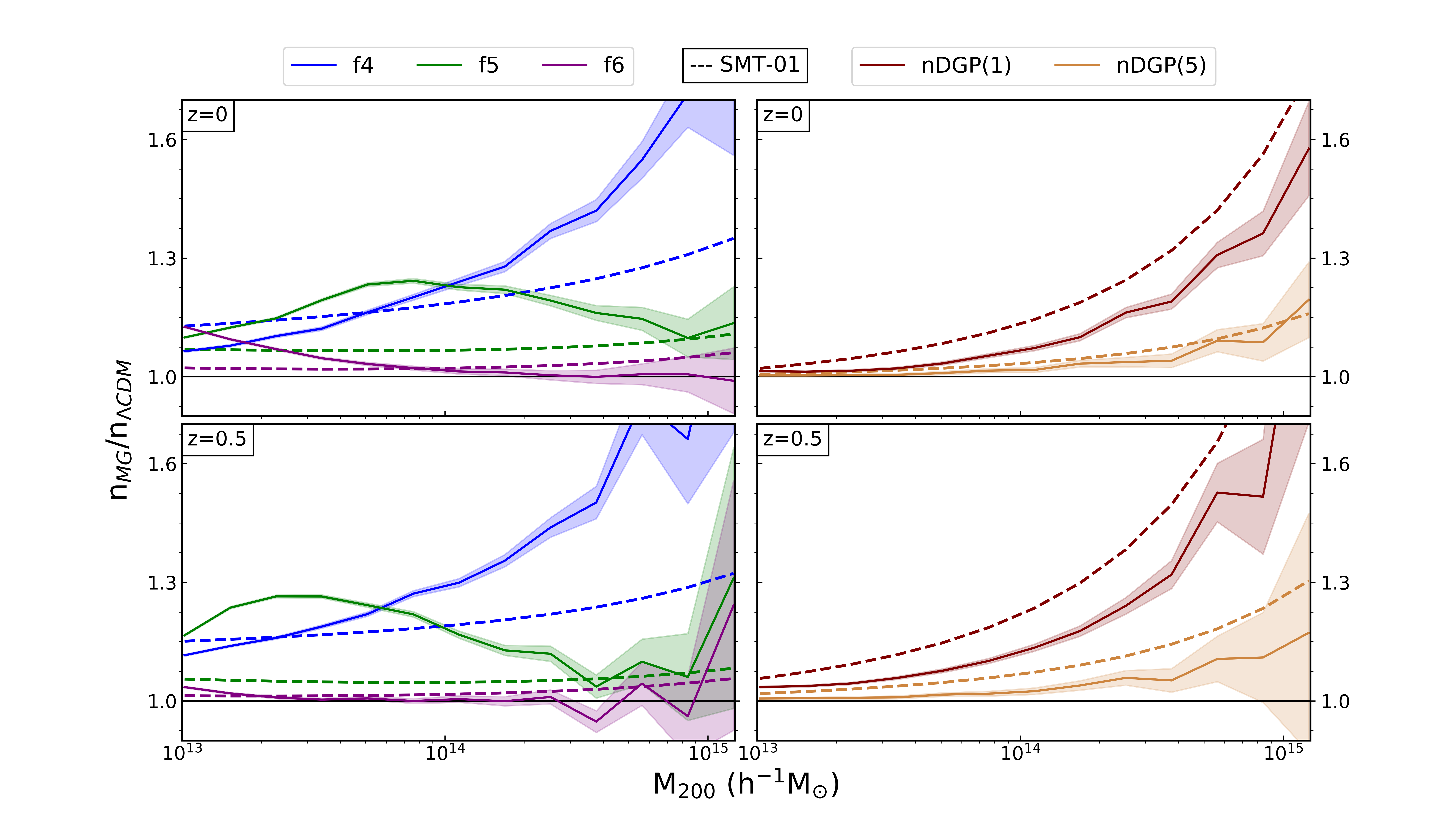}
\caption{Ratio of the differential HMF between MG models and \lcdm, $n_\text{MG}/n_{\Lambda\text{CDM}}$, as a function of halo mass, $M_{200}$, at two redshifts: $z = 0$ (\textit{top panels}) and $0.5$ (\textit{bottom panels}). The MG models considered here are three variants of $f(R)$: $f6, f5$ and $f4$ (\textit{left-side plots}) and two variants of nDGP: nDGP(5) and nDGP(1) (\textit{right-side plots}). The dashed lines mark the ratio of \smt{} predictions \citep{S01_MF} and the solid lines of the same color are the corresponding simulation results. The shaded regions illustrate the propagated Poisson errors from simulations. }
\label{fig:combined_mf_all_z}
\end{figure*}

An equally important ingredient, along with the form of the halo multiplicity function, to obtain an analytical HMF prediction is the linear theory matter density power spectrum. To calculate this quantity, we use a modified version of the \textsc{CAMB} cosmological code \cite{CAMB}, including a module implementing the $f(R)$ and nDGP models \cite{MGCAMB2011JCAP}. We examined the modified $\delta_c$ values adjusted for a specific MG model, as suggested by \cite{sphericalcollapse_braneworld} for the nDGP model, and by \citep{sc_hmf_mg} for the $f(R)$. This however yielded HMF theoretical predictions that also significantly differ from our simulations results and fail to the same extent as the \lcdm{}-based $\delta_c$ theoretical model we examine (see text below and the \cref{fig:combined_mf_all_z}). Thus, for the sake of simplicity and clarity, we will use a standard \lcdm{} spherical collapse based $\delta_c$ values for obtaining all our analytical HMF predictions.

We begin by taking a look at the differences between linear theory matter density variance of our models. In \cref{fig:sigma_allz} we show the ratio of $\sigma(M)$ of a given MG model w.r.t.\ the fiducial \lcdm{} case. The results are expressed as a function of the halo mass scale and are shown for a range of redshifts $0 \leq z \leq 10$. The differences in the density variance are driven by the differences in the shape and amplitude of the linear-theory power spectra of the corresponding models. We show the $f(R)$ and nDGP families in separate panels to emphasize on a clearly scale-dependent nature of $f(R)$ linear matter variance deviation from \lcdm{}.

In \cref{fig:combined_mf_all_z} we compare the analytical \smt{} prediction for the HMF with the results obtained from $N$-body simulations. To reduce the potential impact of %the HMF 
model inaccuracies and cosmological dependencies, we focus on the ratio of the MG to \lcdm{} differential HMFs: $n_\text{MG}/n_{\Lambda\text{CDM}}$. Departures of this ratio from unity mark the deviations from the GR-based structure formation scenario induced by the action of the fifth-force. We consider two epochs as an example: $z=0$ (plots in the top row) and $z=0.5$ (plots in the bottom row). For clarity we compare the $f(R)$ model family (left-hand plots) and the nDGP branch (right-hand plots) separately. In each case, the dashed lines mark the ratio of the \smt{} model predictions obtained using the linear theory power spectra of the relevant models. The solid lines in corresponding colors highlight the results obtained from simulations, and the shaded regions illustrate propagated Poisson errors obtained from the halo number count. We see that the theoretical HMF model provides a rather poor match to the MG simulation results. The trend is that the  empirical predictions for MG models fostering weaker deviations from GR, \ie\ nDGP(5) and $f6$ are closer to the $N$-body results. For all our stronger models, the \smt{} predictions catastrophically fail to capture the real non-linear HMF. However, there are trends present in these mismatches, depending on both the redshift and the model. For the nDGP gravity, the theoretical HMF model over-predicts deviations from \lcdm{} for both epochs, while in the case of the $f(R)$ family, the \smt{} model generally under-predicts the real MG effect. We note that for the mass scales considered, the mismatch is larger for $z=0.5$ than for $z=0$ for all the gravity variants, except for $f6$. 

\cref{fig:combined_mf_all_z} illustrates a clear failure of the HMF modeling to capture the real MG physical effect seen in the abundance of halos. For this exercise, we have found that all of the popular HMF models that we tested    \citep[\eg][]{J01_MF,W06_MF,R07_MF,T08_MF,W13_MF,D16_MF} fail here in a similar fashion as the \smt{} model. This is a clear signal that the HMF models, which are based on the extended Press-Schechter theory, are missing some important parts of the physics of the MG models. One, and probably the most significant missing piece is the screening mechanism (as was previously discussed in   \citep{Schmidt_Oyaizu_2008_3,Schmidt_2009_ndgp}). Both the Vainshtein and Chameleon introduce additional complexities to the structure formation, such as the departure from self-similarity in the halo collapse.
In addition, the screening in $f(R)$ gravity is environmentally dependent, which further alters the $n_\text{MG}/n_{\Lambda\text{CDM}}$ ratio. This can be very well appreciated by observing the peak-like feature for the $f5$ model, in which the mass scale of the peak changes with redshift. The combined effects that we have just listed make the construction of accurate analytical HMF models for MG very challenging \citep{extended_est_fR,sc_hmf_mg,fR_nonlinear_structure,universality_hmf_mg_2018,Lombriser_2013,est_fR,simulations_vainshtein,borderline_f6_hmf}.

\section{Testing the universality of the multiplicity function in modified gravity models}
\label{sec:universality_tests}

\begin{figure*}
\includegraphics[width=0.49\textwidth]{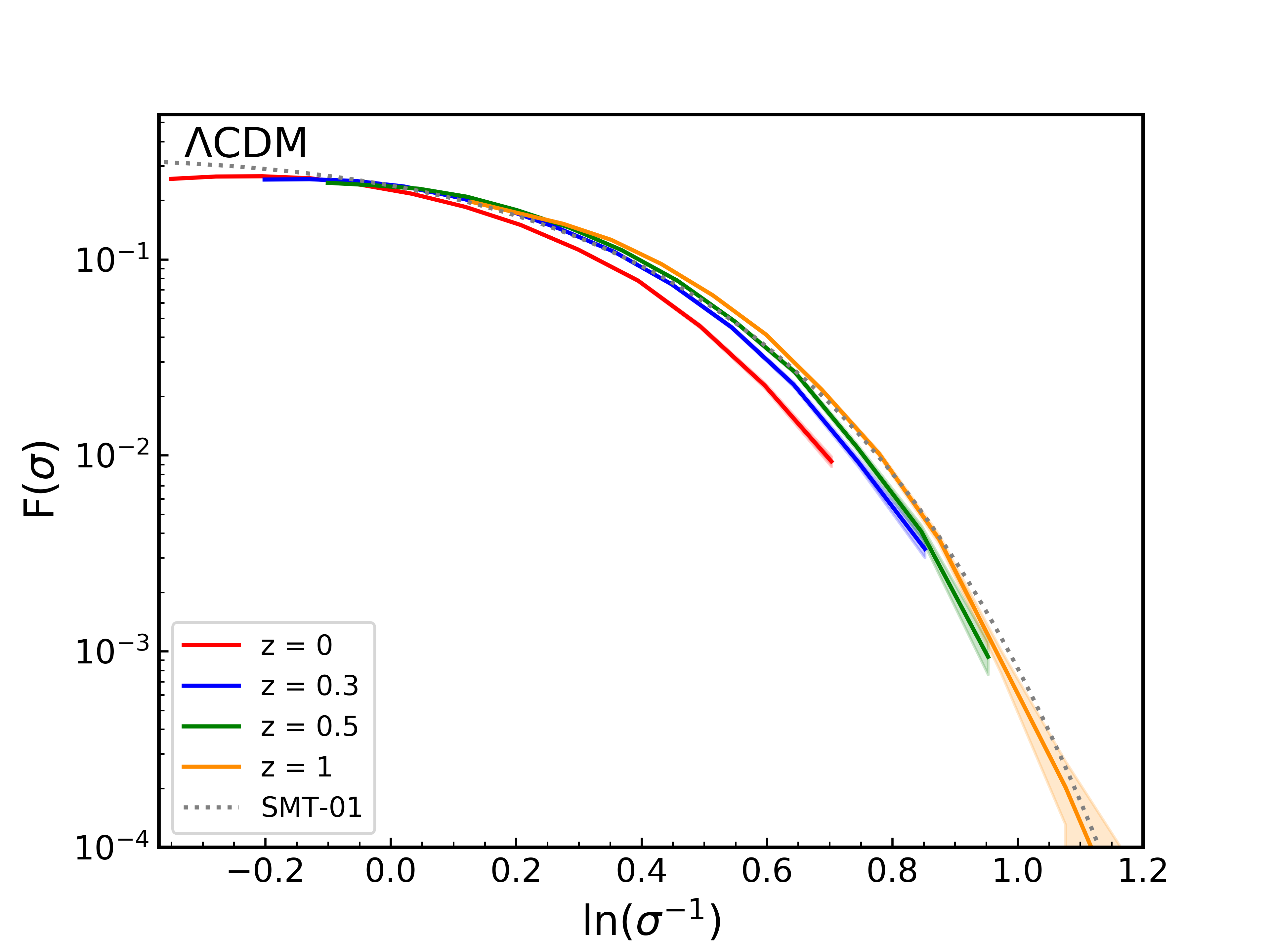}
\includegraphics[width=0.49\textwidth]{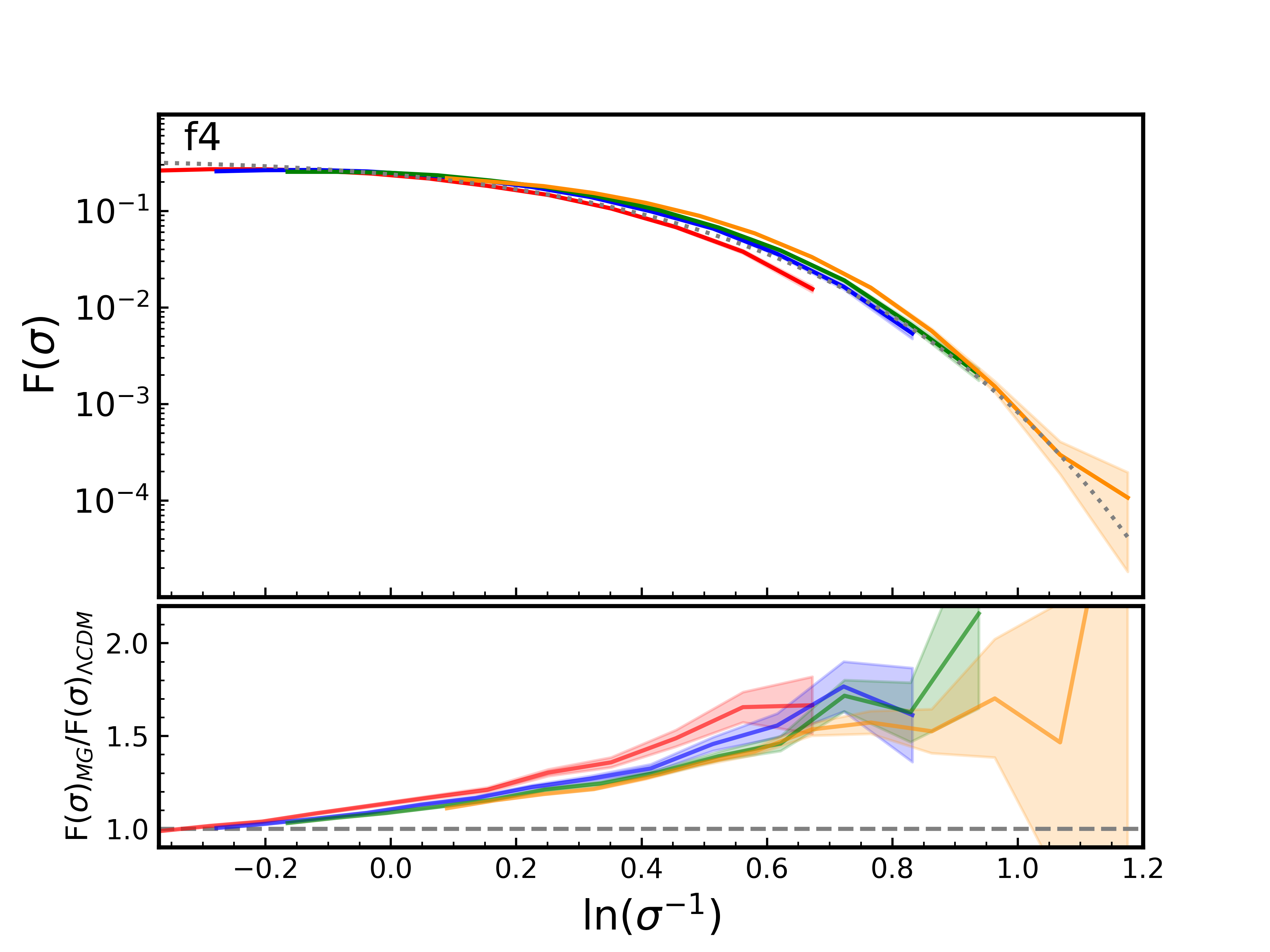}
\includegraphics[width=0.49\textwidth]{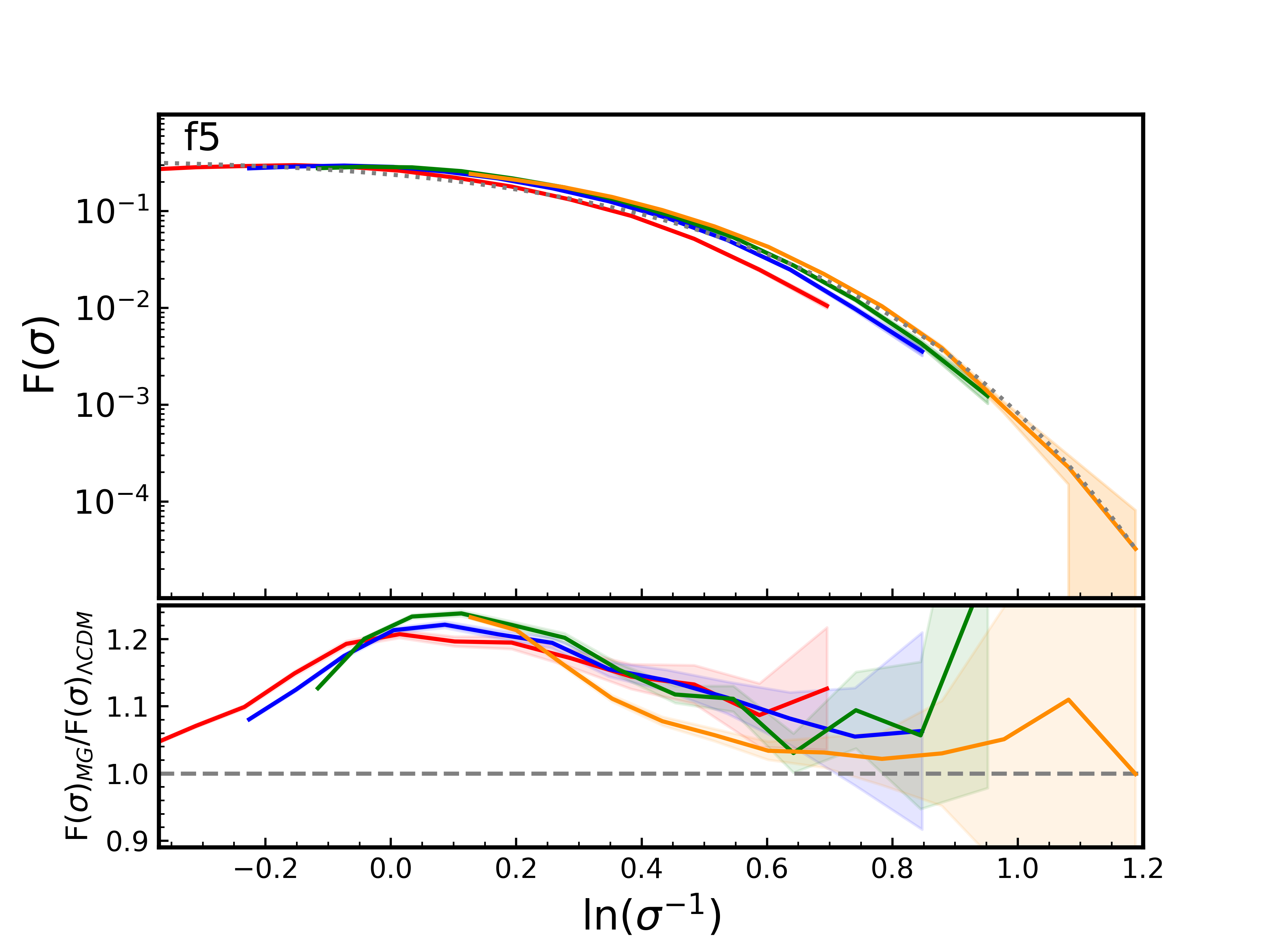}
\includegraphics[width=0.49\textwidth]{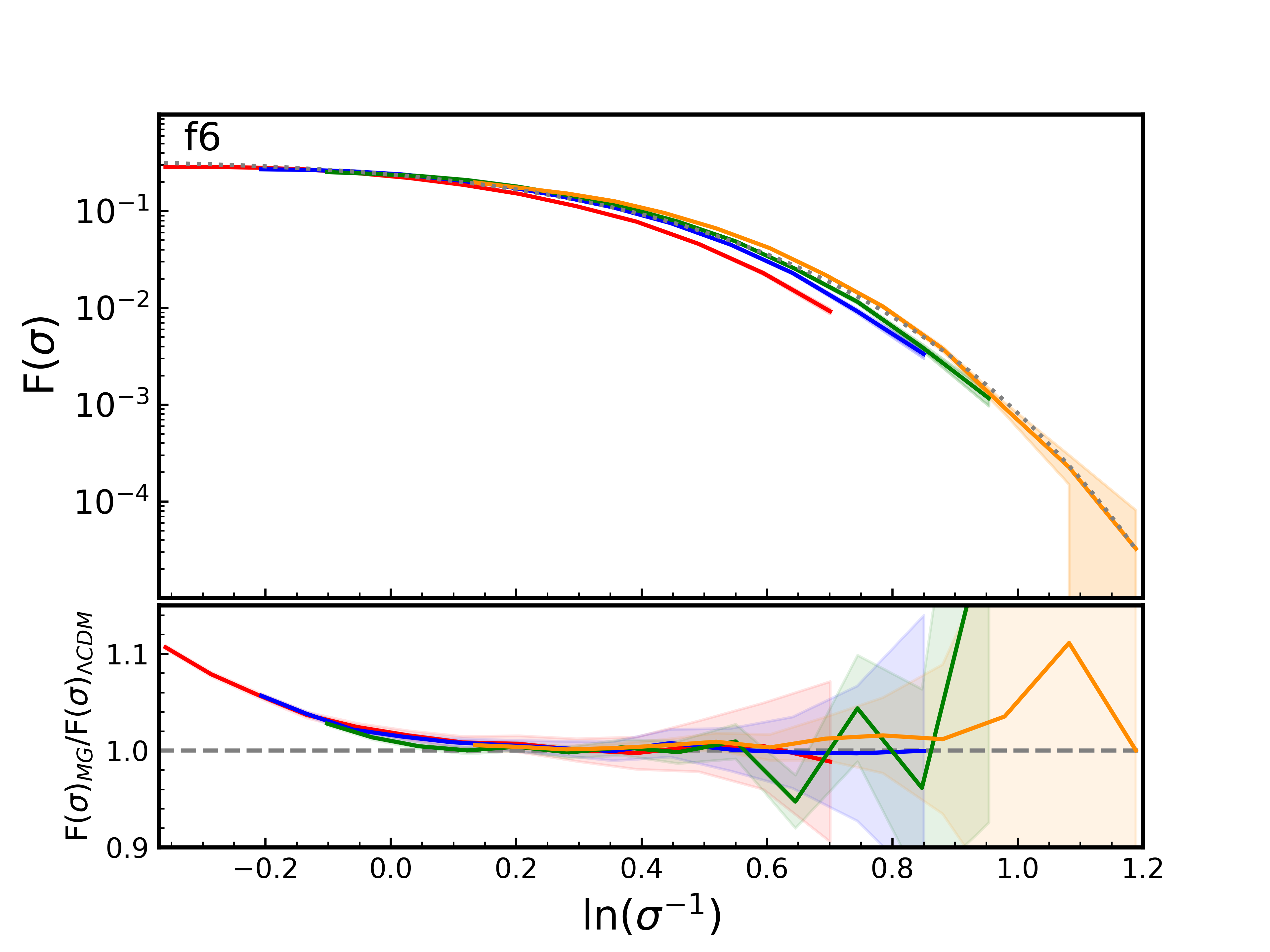}
\includegraphics[width=0.49\textwidth]{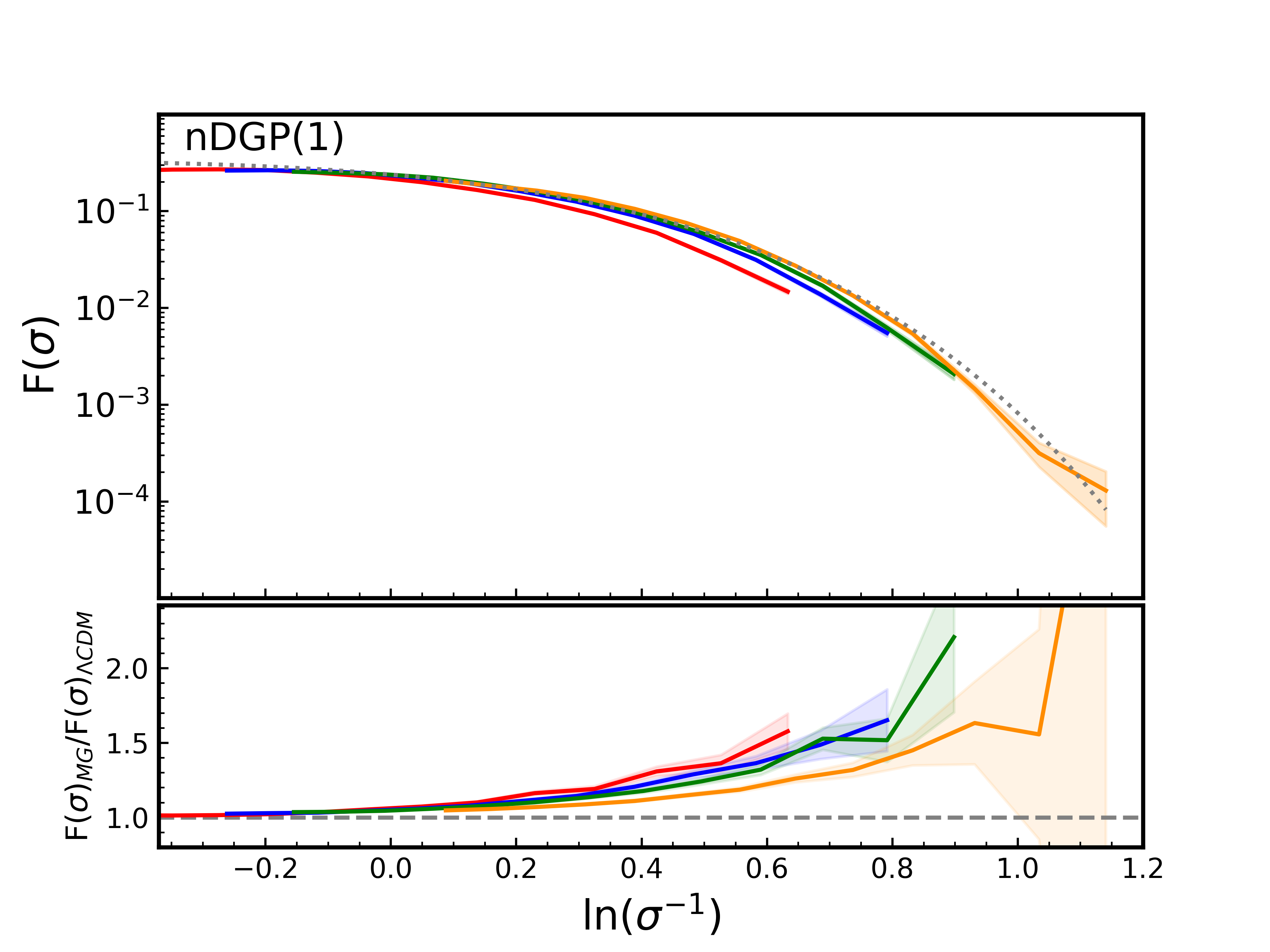}
\includegraphics[width=0.49\textwidth]{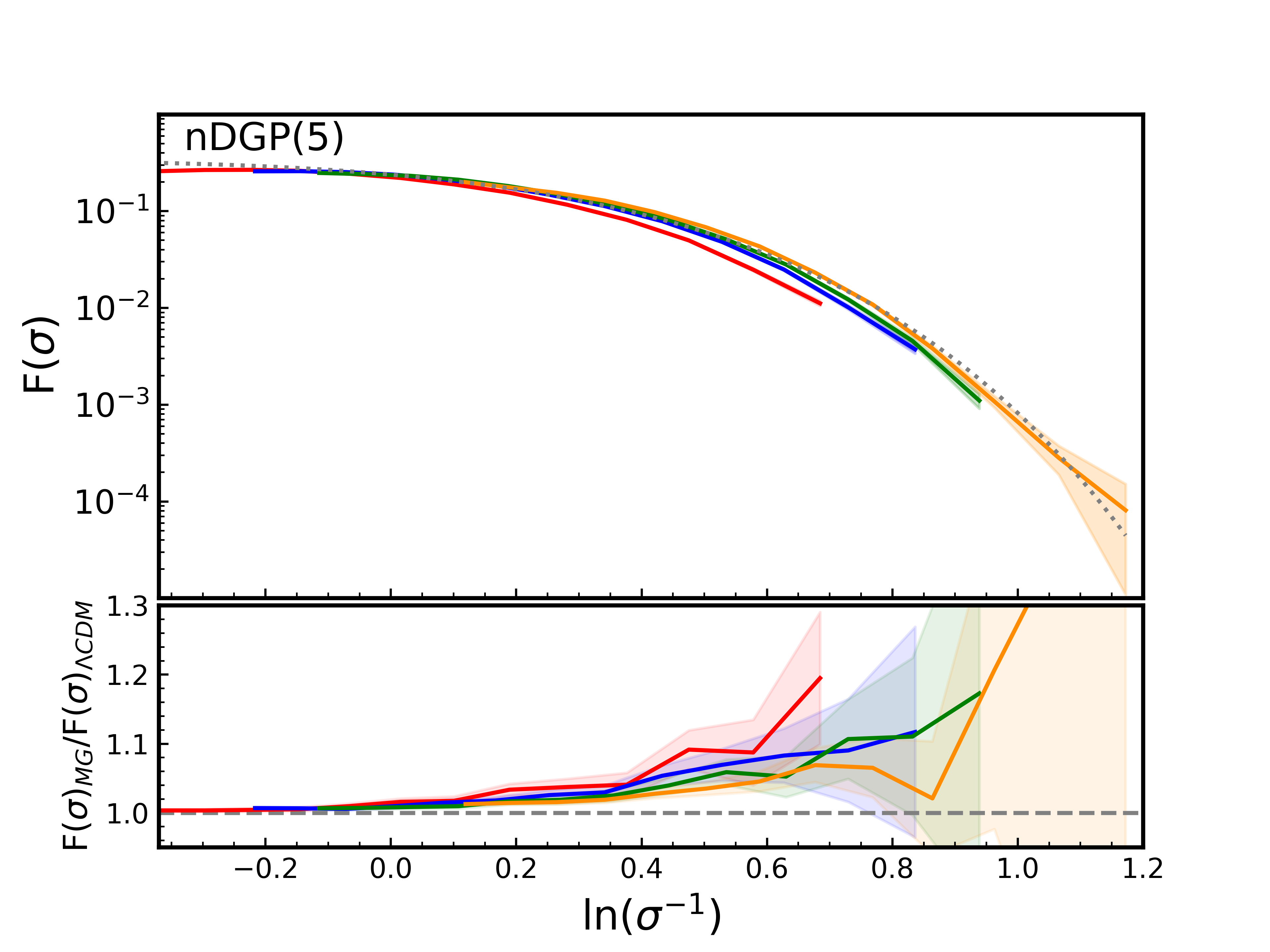}
 \caption{{\it Top sub-panels:} Halo multiplicity function, $\fsig$, for the six gravity models (solid lines with shaded uncertainty regions) as a function of ln($\sigma^{-1}$). Each color corresponds to a different redshift as indicated in the legend in the top-left panel. Dotted grey lines are the analytical \smt\ predictions \cite{S01_MF}. {\it Bottom sub-panels:} Ratio of the MG $\fsig$ to that of \lcdm{} for each redshift.}
  \label{fig:f_allz}
 \end{figure*}

%This first modern formalism for the mass function was first proposed in    \citep{PS74_MF} which gave the first quantitative insight in predicting abundance of regions that collapse to form halos. PS were the first to combine the statistics of the initial density field to the spherical collapse model of top-hat overdensity.

The halo mass and the corresponding scale RMS density fluctuations are connected via a redshift-dependent relation, $\sigma(M,z)$, obtained by plugging \cref{eqn:lagrangian_radius} into  \cref{eqn:SIGMA_EQN}.
When the multiplicity function, $\fsig$, is expressed as a function of ln($\sigma^{-1}$), rather than of the halo mass,   the resulting $\fsig$-ln($\sigma^{-1}$) relation becomes independent of redshift in \lcdm{}    \citep{J01_MF}. This {\it universality} of the halo multiplicity function was shown to hold for various redshifts and  a range of $\sigma^{-1}$   \citep{S01_MF,J01_MF,MF_WHITE,W06_MF,LUKIC_2007,W13_MF,D16_MF}. For the times and scales where this universality holds, one can describe the abundance of structures using only one uniform functional shape of the halo multiplicity function. This approximately universal behavior is a result of the scaling term between $M$ and $\sigma$, $d \ln \sigma/d \ln M$, in \cref{eqn:HMF_MAIN}, which encapsulates the dependencies on the linear background density field evolution, cosmology, and redshift. The universality of the HMF can also be understood as a result of the interplay of two effects in hierarchical cosmologies: firstly, at a fixed mass, enhancement of fluctuations is greater at smaller redshifts, and secondly, a fixed mass would correspond to a larger amplitude of fluctuation at larger $z$ compared to that at a smaller $z$. 

Some of the MG-induced effects will be already encapsulated in the changes of the $\sigma(M)$ relation (illustrated in \cref{fig:sigma_allz}), as shown from both the linear theory
and simulation-based power spectrum studies    \citep{ALAM2020_ELEPHANT,Schmidt_2009_ndgp,simulations_vainshtein,MG_SIGNATURES_HIERARCHICAL_CLUSTERING,Oyaizu_2008_2,hmf_fR_cw,mg_code_comparison,galaxy_formation_braneworld}. Thus, one can hope that when we express HMF using the natural units of the density field fluctuation variance, rather than a specific physical mass, the MG features from \cref{fig:combined_mf_all_z}, which display strong time and scale-dependent variations, will become more regular. This in turn would admit more accurate and straightforward HMF modeling in MG.

We are now interested in studying the halo multiplicity function, $\fsig$, in our MG models. We want to check its behavior and relation w.r.t.\  the standard \lcdm{} case across fluctuation scales, ln($\sigma^{-1}$), and for different epochs. The combined results for all our models are collected in the six plots of \cref{fig:f_allz}. Firstly, let us take a look at the \lcdm{} results, which is shown in the upper-left plot. The colors mark the $\fsig$ computed from snapshots at different redshifts and the shaded regions illustrate the scatter around the mean for different realizations, estimated as Poisson errors from halo number counts (using the definition given in \cite{LUKIC_2007}). With the dotted line, we show the \smt{} model prediction to verify the predicted universal shape of the halo multiplicity function. It is clear that the simulation results already for the vanilla \lcdm{} case are admitting the universality only approximately. Remembering that here $M \propto \sigma^{-1}$, we can say that in the intermediate-mass regime, the agreement between the simulation and the \smt{} formula is the best. For the negative ln($\sigma^{-1}$) regime, the simulations contain a bit fewer halos w.r.t.\ the theoretical prediction and such deficiency is also visible in the high mass (\ie{} small $\sigma$) regime, although there the discrepancy appears somewhat more significant. The $z=0$ case merits a separate comment. Here, the disagreement with the \smt{} formula, and simultaneously also with all the higher redshift simulation data, is very noticeable. The small- and high-mass deficiency of the simulated $\fsig$ is a well-known effect due to the impact of both the discreteness (small masses) and the finite volume (large masses) on the resulting simulated DM density field  \citep{smith_halomodel,hmf_reed_2013,W06_MF, Comparat2017MNRAS, Seppi2021AA}. These effects are combined and especially pronounced for the $z=0$ case, where the density field is most evolved, and thus the most non-linear. %, stage. 
However, given the vast dynamical scale both in $\fsig$ and $\sigma^{-1}$, the regular and approximately universal results for the \lcdm{} run are very encouraging. 

Moving to the MG HMF, we present each variant of the models separately. In the top sub-panel of each MG plot, we observe that the $\fsig$ trend is very similar to the one observed in \lcdm, as discussed above. What seems more interesting, though, are the lower sub-panels of these plots, where we show the relative difference taken w.r.t.\ the \lcdm{} case at each redshift consequently. Each model exhibits a unique and specific combination of both amplitude and scale of departure from the fiducial GR-case. For both $f4$ and nDGP(1), the excess of $\fsig$ can reach up to $\sim 60\%$ or greater in the rare density fluctuation regime, ln($\sigma^{-1})\geq 0.7$. At the same time, both mild variants, $f6$, and nDGP(5) do not depart from GR by more than $\sim 15-20\%$. The departure of  $f5$ from \lcdm{}, which is $\sim 20-25\%$ max, lies between the results of $f4$ and $f6$, as expected. 

The most striking and important observation is the much more enhanced regularity of $\fsig$ departures from the \lcdm{} case when compared with the previous plot of the HMF itself (\cref{fig:combined_mf_all_z}), where the abundance of objects was shown as a function of their mass. Now, when we express $\fsig$ in its natural dimensionless units of the density field variance, ln($\sigma^{-1}$), the MG effects are much more regular across the redshifts.

%Here
However, we see an exception for the $f4$ gravity case in which there are clear signs of deviation from universality, especially at $z=0$. We attribute this to the enhanced fifth force in this $f(R)$ variant which accumulates as time elapses, and has a maximum effect at later redshifts. We note, however, that this rather extreme MG model is unlikely to be valid taking into account the current observational constraints   \citep{HS_fR_2007,fR_constraint_abundance}. Nevertheless, we have considered $f4$ in our analysis for completeness.

In general, the $\fsig$ modification for the $f(R)$ variants follows a very similar pattern as a function of ln($\sigma^{-1}$) and the universality of $\fsig$ over the redshifts is largely established. In addition, the shape of the $\fsig$ modification displays a peak for $f5$, monotonically increases for $f4$, and monotonically decreases for $f6$. This is a clear manifestation of a complicated and non-linear interplay between the local density and the Chameleon screening efficiency. The interpretation could be that the Chameleon mechanism is self-tuned by the environment-dependent non-linear density evolution. As a result, as clearly shown for the $f(R)$ plots of \cref{fig:f_allz}, the self-similarity of the HMF and density field evolution is restored.

The nDGP case is less clear to interpret. The $\sigma(M)$ re-scaling brings the $\fsig$ at all the different redshifts much closer together when we compare with nDGP plots for the HMF in \cref{fig:combined_mf_all_z}. Some resonant time-dependent evolution can be however still noticed; this can be appreciated especially in the case of nDGP(1). This residual redshift-dependence reflects the fact that the screening mechanism in this model family is the Vainshtein, which does not depend on the local density field (\ie{} the environment). Thus, the general force enhancement factor and the resulting growth rate of structures is only time-dependent but scale-independent (as also seen in the right plot of \cref{fig:sigma_allz}), which forbids the self-tuning mechanism in nDGP, unlike the case of $f(R)$. %, as discussed in the previous paragraph. 
However, the nDGP force enhancement factor, $\Xi(z)$ can be easily calculated for any given redshift    \cite{nonlinear_interactions_nDGP}. We exploit this to introduce time-dependent physical re-scaling for nDGP, $\widetilde{\sigma} \equiv \sigma/\Xi(z)$ which has been defined and discussed in the Appendix \ref{appendix:rescale_nDGP}. This physically-tuned re-scaled factor removes nearly all the redshift dependence in the nDGP $\fsig$ modification (see also \cref{fig:n1_scaling_noscaling}). Thus, this one additional step can restore the expected self-similar and universal behavior of the HMF in the normal branch of the Dvali-Gabadadze-Porrati model.

%%%%%%%%%%%%%%%%%%%%%%%%%%%%%%%%%%%%%%%%%%%%%%%%%%%%%%%%%%%%%%%%%%%%%%%%%%%%%%%%%%%%%%%%%%%%%%%%%%%%%%%%%%%%%%%%%%%%%%%%%%%%%%%%%%

\section{Re-scaling the multiplicity function in modified gravity}
\label{sec:fsigma_mg}

We will now exploit the universality of the MG halo multiplicity function, which was established in the previous section, to characterize the essential effects induced by beyond-GR dynamics on the abundance of halos. We do this by first finding an enclosed formula that describes well the shape of the MG HMF departure from the fiducial \lcdm{} case. Then, we obtain the best fit for a given MG-variant for all redshifts. Finally, we test our newly found MG HMF model against data sets that were not used for the fitting. In this section, we will cover the first two steps, leaving the testing for \S\ref{sec:test_fits}.

As we have already highlighted, $\fsig$ displays some degree of universality for each of the gravity models in a sense that it takes approximately the same form for all the considered redshifts (as shown in \cref{fig:f_allz}). This has already been studied extensively for \lcdm{}, taking advantage of which many fitting functions have been proposed to formulate the \lcdm{} HMF   \citep[e.g.][]{J01_MF,S01_MF,W13_MF,D16_MF,W06_MF}. However, some other authors have in contrast reported a departure from universality in \lcdm{}. This has been identified as the dependence of HMF on several factors: redshift and cosmology   \citep{R07_MF,T08_MF,C10_MF,C11_MF,MALLEA2021_MF}; non-linear dynamics associated with structure formation \cite{C11_MF,D16_MF}; some artificially induced factors like the type of mass definition employed to define a halo \citep{mass_halo_white,MF_WHITE,T08_MF,W13_MF,splashback_universal_mf,D16_MF}; the value of the linking length \cite{more_linking_length_mf,C11_MF}; or numerical artifacts in the simulations and computations of the HMF \citep{LUKIC_2007,R07_MF}. For the case of MG models like $f(R)$ and nDGP, the universal trend in the HMF is even less certain, given the explicit dependence of the fifth force and of the associated screening mechanisms on the scale and environment    \citep{extended_est_fR,Lombriser_2013,fR_nonlinear_structure,universality_hmf_mg_2018,dynamical_mass_mg,Schmidt_2009_ndgp}. 

Considering the failure of the empirical relations devised for the \lcdm{} HMF to capture MG effects (as discussed in \S\ref{sec:modeling_mf}), rather than searching for a universal MG $\fsig$ fit, we adopt a different approach. We instead characterize the beyond-GR HMF as a functional deviation from the \lcdm{}-case. Thus, we express the targeted MG HMF as a function of the \lcdm{} case at a fixed $\sigma$, \ie{} $F_\text{MG}[F(\sigma)_{\Lambda\text{CDM}}]$, and the general form of the relation between \lcdm{} and MG multiplicity functions is given by:
\begin{equation}
    \label{eqn:mul_fit_eqn}
    \fsig_\text{MG} = \Delta_{\text{MG}} \times \fsig_{\Lambda \text{CDM}}.
\end{equation}
Here $\fsig_{\Lambda \text{CDM}}$ is obtained from \lcdm{} simulations. As we are using $N$-body derivations to characterize $\fsig$, the resulting MG HMF model will automatically incorporate non-linear effects to the limit of our simulations.

The general conclusion from \S\ref{sec:universality_tests} is that the ratio $\fsig_\text{MG}/\fsig_{\Lambda\text{CDM}}$ has an approximately universal, redshift-independent shape after re-scaling the fluctuation scales in terms of ln($\sigma^{-1}$). We calibrate the relation \eqref{eqn:mul_fit_eqn} based on the \elephant{} data and find enclosed formulas to capture $\Delta_{\text{MG}}$ as probed by the simulations for each of the MG variants we have considered.

%%%%%%%%%%%%%%%%%%%%%%%%%%%%%%%%%%%%%%%%%%%%%%%
\subsubsection{$f(R)$ gravity} 
\begin{table}
\caption{Parameters for the $f(R)$ gravity model fit, $\Delta_{f(R)}$ (Eq.\ \ref{eqn:fit_fR_equation}).}
% 4 significant figures
\label{table:fit_fR_param}

\begin{tabular}{|c|c|c|c|}
\hline
Model & $a$ & $b$ & $c$  \\
\hline
\hline
$f4$ & 0.630 & 1.062 & 0.762 \\
\hline
$f5$ & 0.230 & 0.100 & 0.360  \\
\hline
$f6$ & 0.152 & -0.583 &  0.375 \\
\hline
\hline
\end{tabular}
\end{table}
We have found that the following analytical expression can be used to fit $\Delta_{\text{MG}}$ in $f(R)$ simulations:
\begin{equation}
 \Delta_{\text{MG}} \equiv \Delta_{f(R)} = 1 + a \exp \left[-\frac{(X-b)^2}{c^2}\right],
% \, e^{-(\text{X}-b)^{2}/c^{2}}\,\,,  
\label{eqn:fit_fR_equation}
\end{equation}
where $X \equiv$ ln($\sigma^{-1}$). Here, the parameter `$a$' sets the maximum value of $\fsig_{f(R)}/\fsig_{\Lambda \text{CDM}}$, `$b$' corresponds to the value of ln($\sigma^{-1}$) at the maximum enhancement and '$c$' determines the range of ln($\sigma^{-1}$) across which  $\fsig_{f(R)}$ is enhanced w.r.t.\  $\fsig_{\Lambda \text{CDM}}$.
These best-fit parameters were obtained by solving for the minimum reduced $\chi^{2}$, which were procured by comparing the ratio $\fsig_{f(R)}/\fsig_{\Lambda \text{CDM}}$ from simulations with our analytical expression \eqref{eqn:fit_fR_equation} across all redshifts and for each ln$(\sigma^{-1})$ bin. For $f(R)$, the best-fit values of the parameters are given in \cref{table:fit_fR_param}. 

\begin{figure}
\includegraphics[width=\columnwidth]{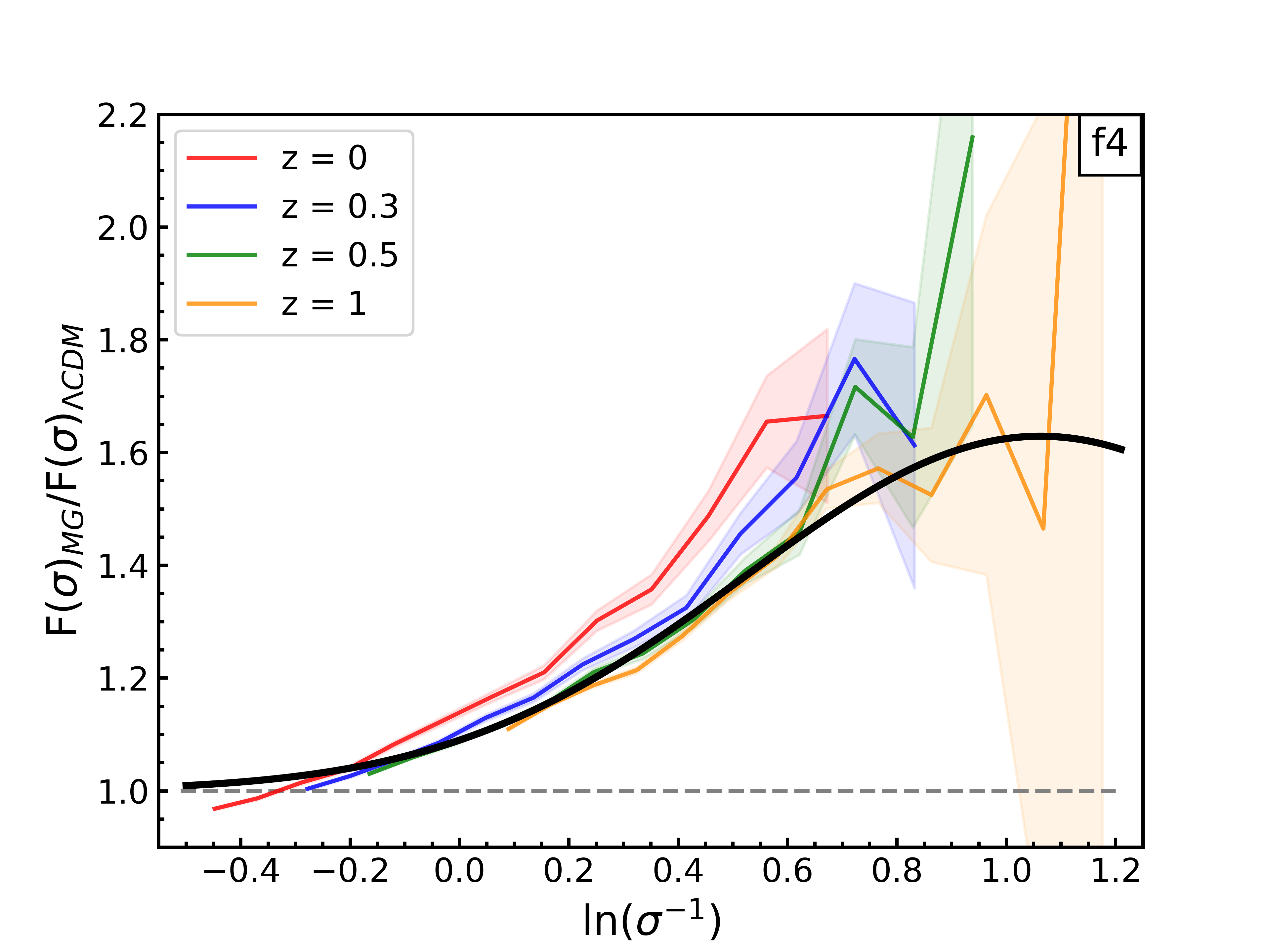}
\includegraphics[width=\columnwidth]{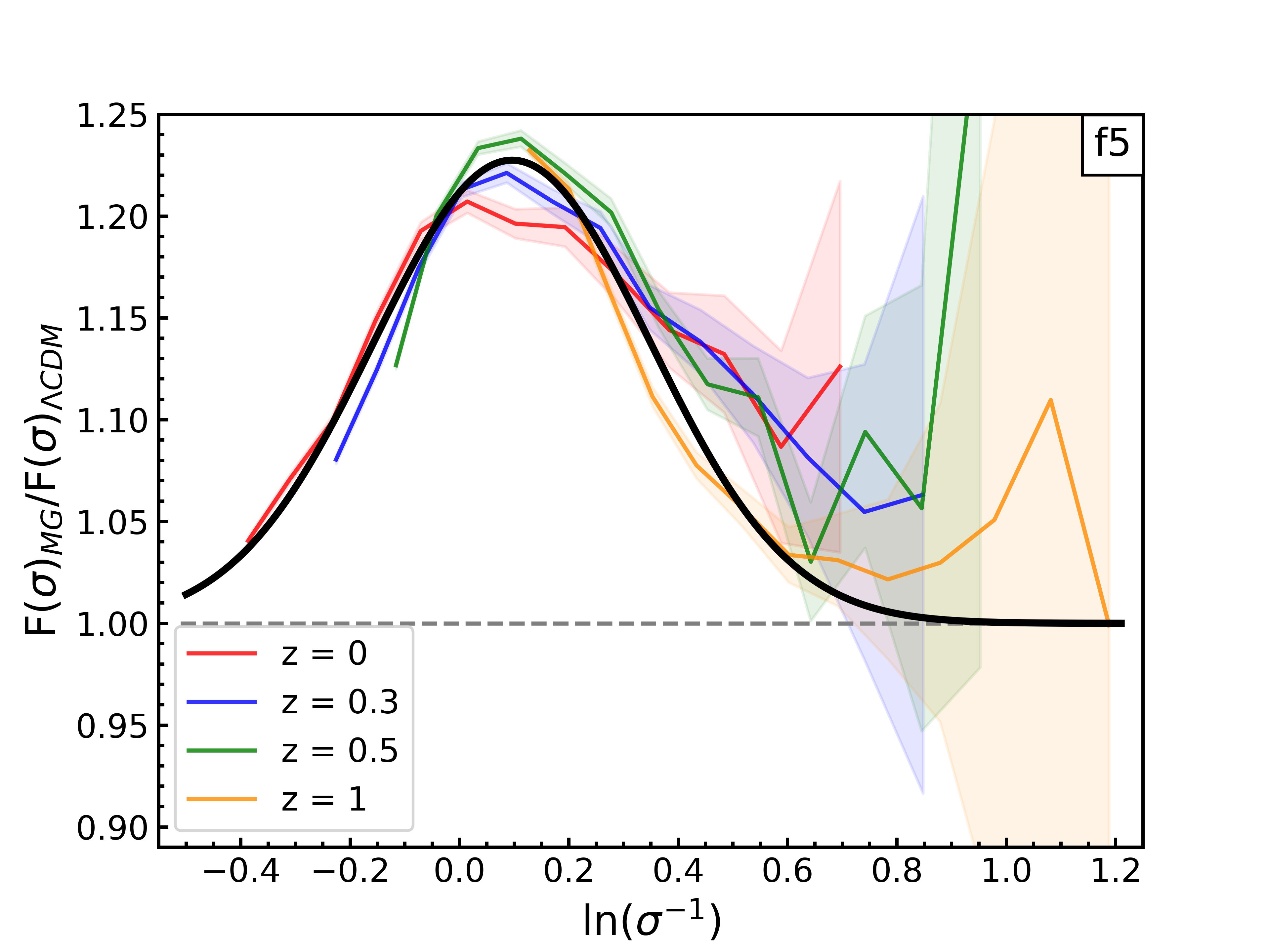}
\includegraphics[width=\columnwidth]{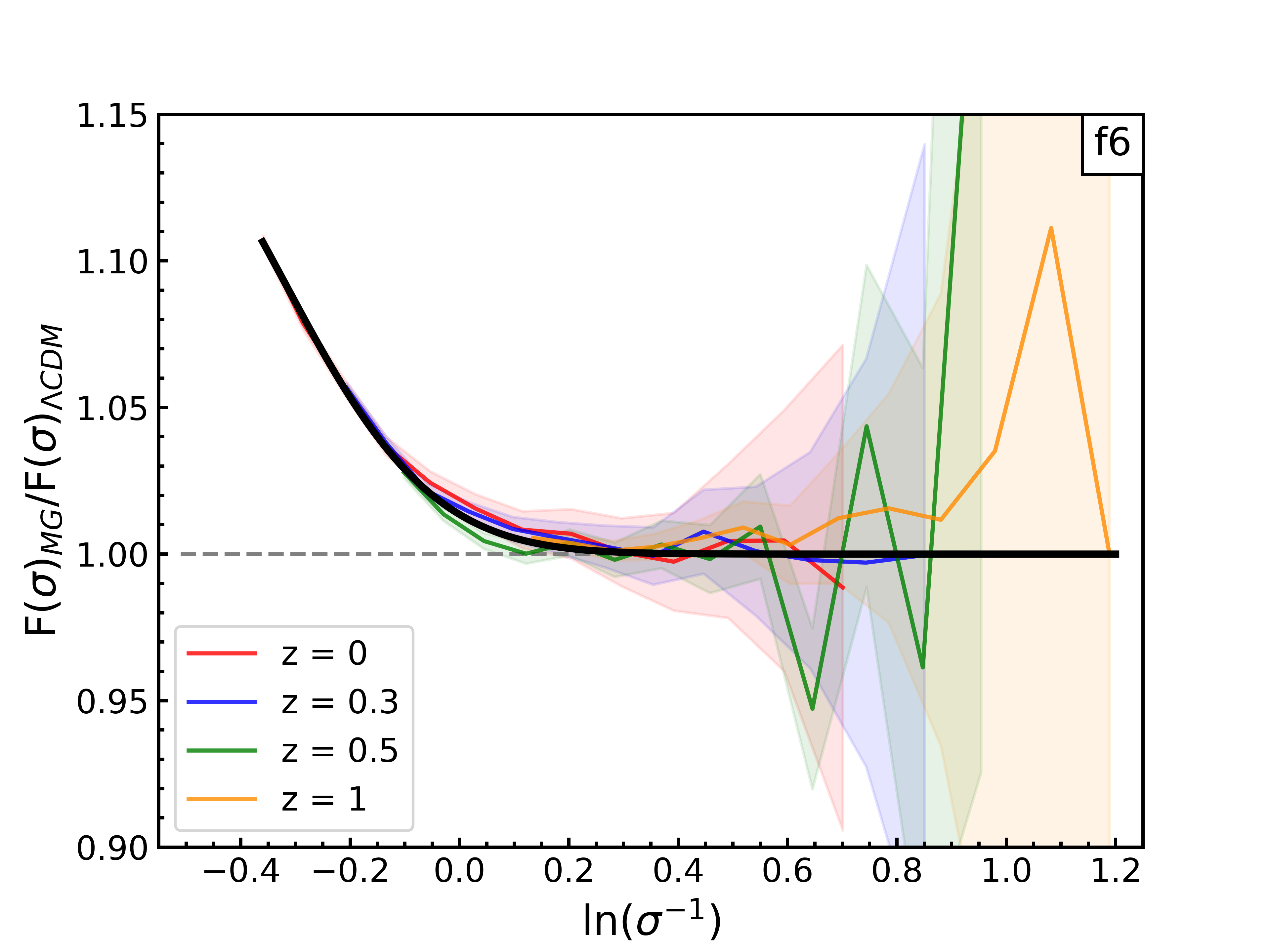}
 \caption{Ratio of the halo multiplicity functions $\Delta_{f(R)} = {\fsig_{f(R)}}/{\fsig_{\Lambda\text{CDM}}}$ for $f4$, $f5$ and $f6$. Colored lines indicate the four different redshifts. The black line in each panel is our universal fit (Eq.\ \ref{eqn:fit_fR_equation}) with the best-fit parameters provided for each model in \cref{table:fit_fR_param}. Poisson errors in both $\fsig_{f(R)}$ and $\fsig_{\Lambda\text{CDM}}$  are propagated to plot the error ranges of the ratio.}
\label{fig:fit_fR_plots}
\end{figure}

In \cref{fig:fit_fR_plots}, we illustrate the performance of our analytical formula with the best-fit parameters from \cref{table:fit_fR_param} by comparing it with the simulation data for each redshift. As we can observe, generally the accuracy of our formula is good at high-$\sigma$ (or low-mass) regime, where also the lines indicating  different redshifts are close to each other, which is not surprising given the good statistics of our simulation in this regime. At the low-$\sigma$ (high-mass and rare-object) regime, the data is characterized by a much bigger scatter. This drives our best fit sometimes in between the different redshift lines, an effect most visible for the $f5$ case. As the fit was obtained together for all the redshifts, and the assumed universality holds only approximately, we expect that the deviation at some redshifts might be larger compared to others. Nonetheless, given the scatter of both the mean trends and their corresponding errors, our fitting formula does a remarkably good job.

The general Gaussian form of \cref{eqn:fit_fR_equation} fosters a `peak-like' feature with some specific ln($\sigma^{-1}) = b$ value for the peak location. This suggests that the HMF of $f(R)$-gravity models can be characterized by a new universal scale: a scale at which the combined effect of the enhanced structure formation and ineffective screening mechanism maximizes the halo abundance for the case of $f(R)$ gravity.

Given the limitations of the \elephant{} simulations, we can expect that our best-fit parameters could be probably still tuned even more, if bigger and higher-resolution simulations became available. Nonetheless, considering these limits, we appreciate that the resulting reduced $\chi^2$ values of the best-fits, both for individual redshifts, as well as for the concatenated redshift data, are very reasonable. These reduced $\chi^2$ values were obtained by comparing the MG HMF obtained using simulations and our \cref{eqn:mul_fit_eqn}, with parameters taken from \cref{table:fit_fR_param}. We give them in \cref{table:chisqr_elephant}.

\subsubsection{nDGP gravity}
\label{subsub:fit_nDGP}
\begin{table}
\caption{Parameters for the nDGP gravity model fit, $\Delta_{\text{nDGP}}$ (Eq.\ \ref{eqn:fit_nDGP_equation}).}
    \label{table:fit_nDGP_param}
\begin{tabular}{|c|c|c|c|c|}
\hline
Model & $p$ & $q$ & $r$ & $s$\\
\hline
\hline
nDGP(1) & 1.35 & 0.258 & 5.12 & 4.05    \\
\hline
nDGP(5) & 1.06 &  0.0470 & 11.8 &  4.19   \\
\hline
\hline
\end{tabular}
\end{table}

\begin{figure}
    \includegraphics[width=\columnwidth]{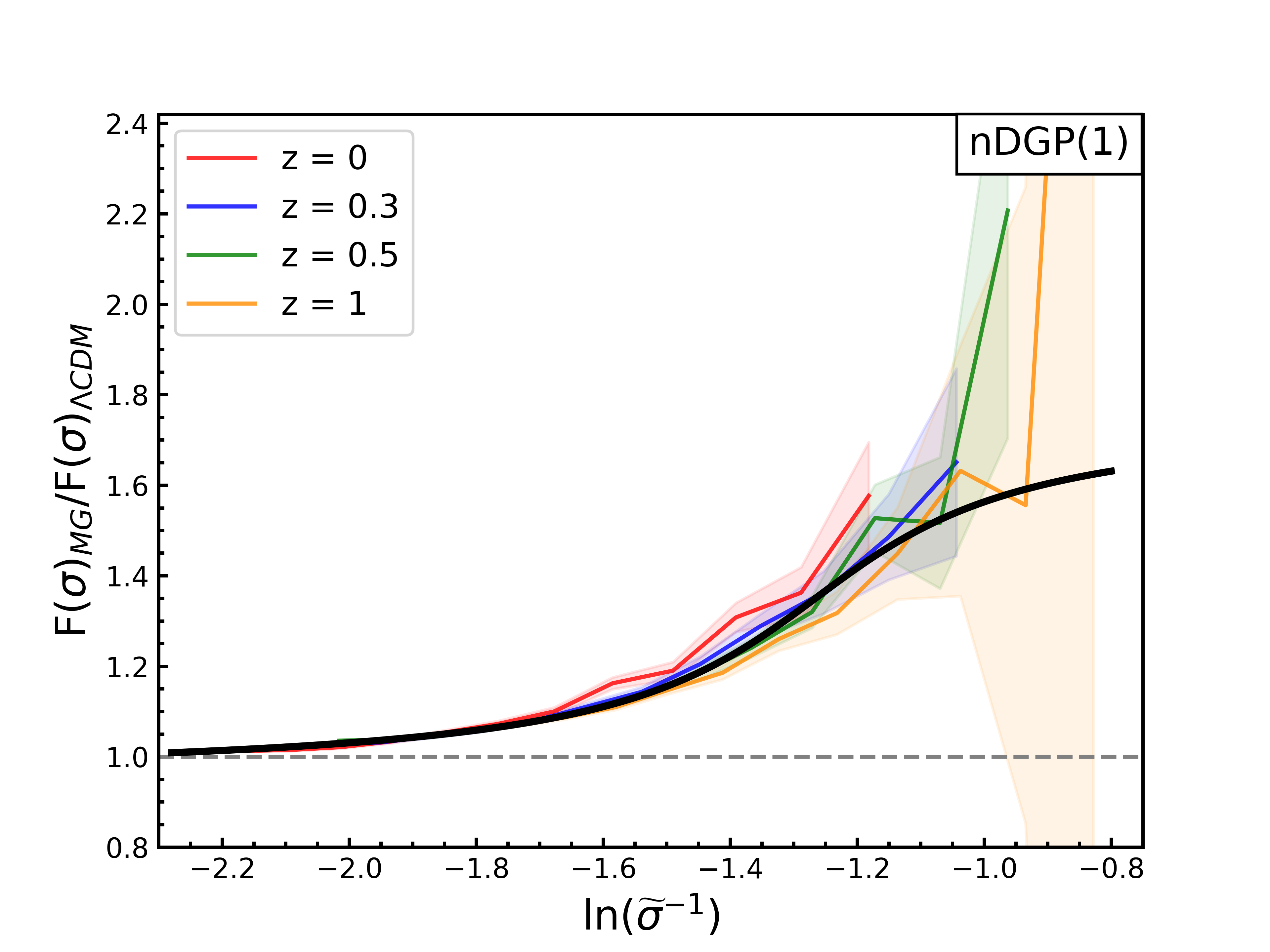}
    \includegraphics[width=\columnwidth]{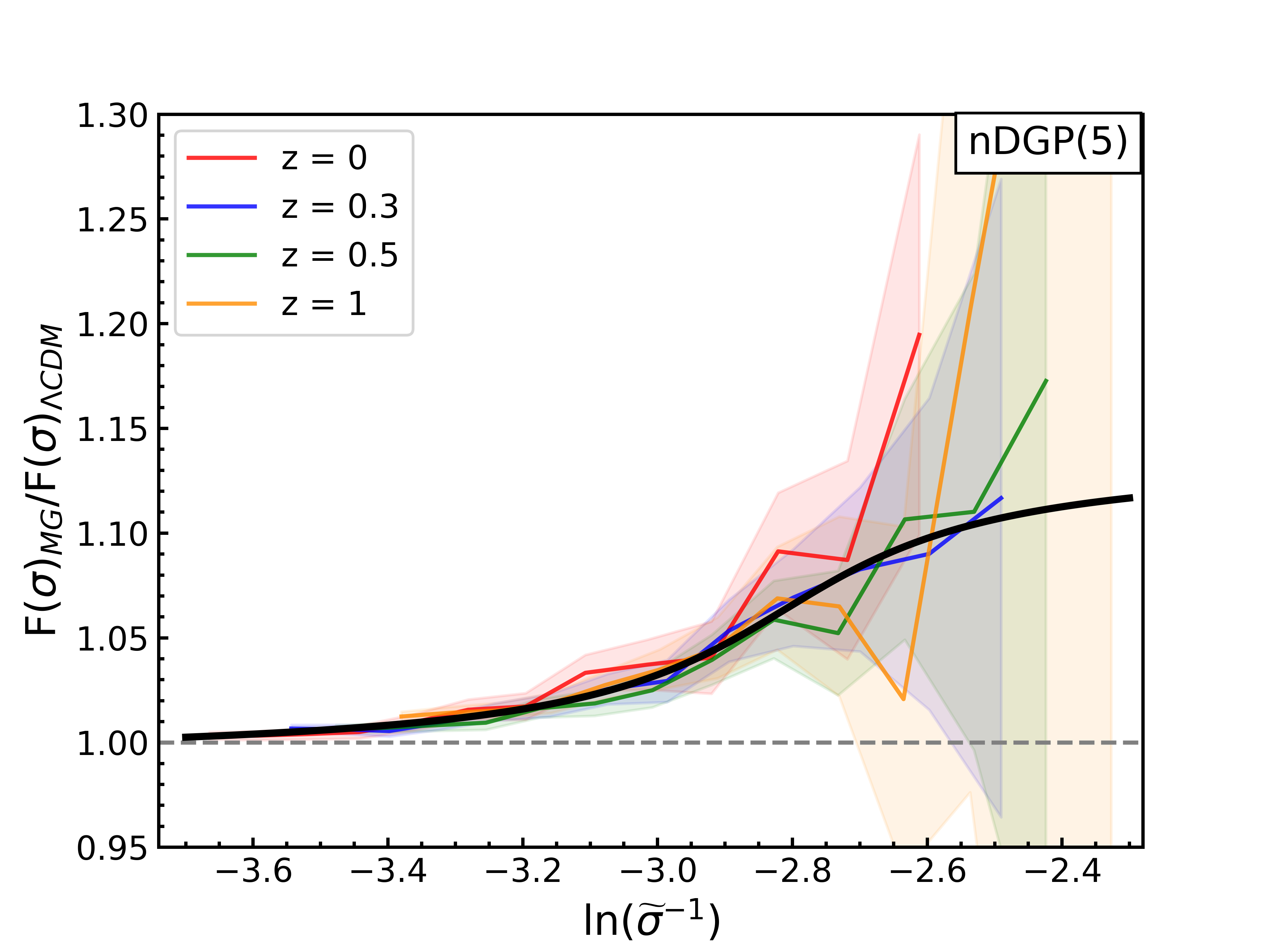}
\caption{Ratio of the halo multiplicity functions  $\Delta_\text{nDGP} = {\fsig_\text{nDGP}}/{\fsig_{\Lambda\text{CDM}}}$ for nDGP(1) and nDGP(5).  Colored lines indicate the four different redshifts. The black line in each panel is our universal fit (Eq.\ \ref{eqn:fit_nDGP_equation}) with the best-fit parameters provided for each model in \cref{table:fit_nDGP_param}. Poisson errors in both $\fsig_\text{nDGP}$ and $\fsig_{\Lambda\text{CDM}}$  are propagated to plot the error ranges of the ratio. }

\label{fig:fit_nDGP_plots}
\end{figure}

For this class of models, we found a different shape of $\Delta_\text{MG}$, as the formula that works for $f(R)$ failed to provide a good fit to the data. Instead, we use an arctan parametrization that much better captures the nDGP shape of $\Delta_\text{MG}$, given by:
\begin{equation}
   \Delta_{\text{MG}} \equiv \Delta_{n\text{DGP}} = p + q \arctan{(s \, X+r)}.
\label{eqn:fit_nDGP_equation}
\end{equation}
Here, $X$ is the \textit{re-scaled} mass density variance, $X \equiv \ln(\widetilde{\sigma}^{-1})$, and we recall that $\widetilde{\sigma} = \sigma/\Xi(z)$. %In \cref{eqn:fit_nDGP_equation}, t
The parameter '$p$' shifts the lower asymptote of the curve, '$q$' sets the amplitude of $\fsig_{n\text{DGP}}/\fsig_{\Lambda \text{CDM}}$, '$r$' dictates the range of $\ln (\tilde{\sigma}^{-1})$ and '$s$'  determines the slope of the deviation curve.
These best-fit parameters were obtained using the method analogous to the one discussed for $f(R)$ and are given in \cref{table:fit_nDGP_param}. Also,   we plot in \cref{fig:fit_nDGP_plots} the resulting best-fit $\arctan{}$ curves alongside the simulation data at various redshifts. A quick look at the reduced-$\chi^2$ values in \cref{table:chisqr_elephant} indicates that our nDGP fits on average characterize the data even better than it was for the case of $f(R)$ fits. 

\begin{table}
\caption{Reduced $\chi^{2}$ values obtained by comparing our fit equations (\cref{eqn:fit_fR_equation} and \cref{eqn:fit_nDGP_equation}) with the respective simulation results. We show values for individual redshift and for the joint all-z data. }
\label{table:chisqr_elephant}
\begin{tabular}{|c|c|c|c|c|c|}
\hline
Model & $z=0$ & $z=0.3$ & $z=0.5$ & $z=1$ & All $z$\\
\hline
\hline
$f4$ & 18.9 & 6.55 & 1.46 & 4.45 & 7.60\\
\hline
$f5$ & 4.95 & 6.96 & 6.11 & 4.30 & 5.58 \\
\hline
$f6$ & 0.300 & 0.236 & 1.04 & 0.290 & 0.470 \\
\hline
nDGP(1) & 3.66 & 0.834 & 1.23 & 0.860 & 1.64  \\
\hline 
nDGP(5) & 0.364 & 0.203 & 0.332 & 0.428 & 0.385  \\
\hline
\hline
\end{tabular}
\end{table}

%%%%%%%%%%%%%%%%%%%%%%%%%%%%%%%%%%%%%%%%%%%%%%%%%%%%%%%%%%%%%%%%%%%%%%%%%%%%%%%%%%%%%%%%%%%

\section{Testing the fits}
\label{sec:test_fits}

We want to test our best-fits obtained for various MG models with simulation data that were not used for the original fitting. This test will inform us of the degree of both applicability and accuracy of our HMF modeling.

However, as we did not have access to $f(R)$ simulations other than \elephant{}, we limit this cross-check to two nDGP(1) runs.
These will be our test-beds that have better resolution than the data that we used to derive the scaling relations from the previous section. The description of these two independent runs has been given in \S\ref{sec:mg_mod_sims}.

We note that the differences between \elephant\ and the independent simulations that we use here may lead to some complications in the comparisons. We expect that the effects from the different cosmologies will be secondary, as long as we compare the ratios, $\Delta_\text{MG}$, rather than the absolute HMF values. This is because the expansion history and power-spectrum normalization will be the same for \lcdm-Planck15 and nDGP(1)-Planck15, so to the first order, the value of $\Delta_\text{MG}$ will be driven mostly by the fifth-force induced effects. Nonetheless, the re-scaling from $\sigma$ to $\tilde{\sigma}$ in nDGP is cosmology-dependent and could contribute to possible discrepancies (see Appendix \ref{appendix:rescale_nDGP}).

The most valuable aspect of this exercise is that we shall test our best-fit $\Delta_\text{MG}$ models on simulation data that cover different $\sigma(M)$ range than the original \elephant{} suite. The halo masses at various redshifts in the \mgmil{} run are contained within $0.50\lesssim \sigma(M) \lesssim 4.1$, while for \dudeg{} this range is roughly $0.27 \lesssim \sigma(M) \lesssim 1.9$, whereas for \elephant\, the values lie between $0.32 \lesssim \sigma(M)\lesssim 1.5$. Thus, the mass variance of the smaller box reaches to nearly 3 times higher $\sigma(M)$, while of the bigger box to $\sim 40\%$ higher $\sigma(M)$ value, than in the \elephant{} suite.

\begin{figure} 
 \includegraphics[width=\columnwidth]{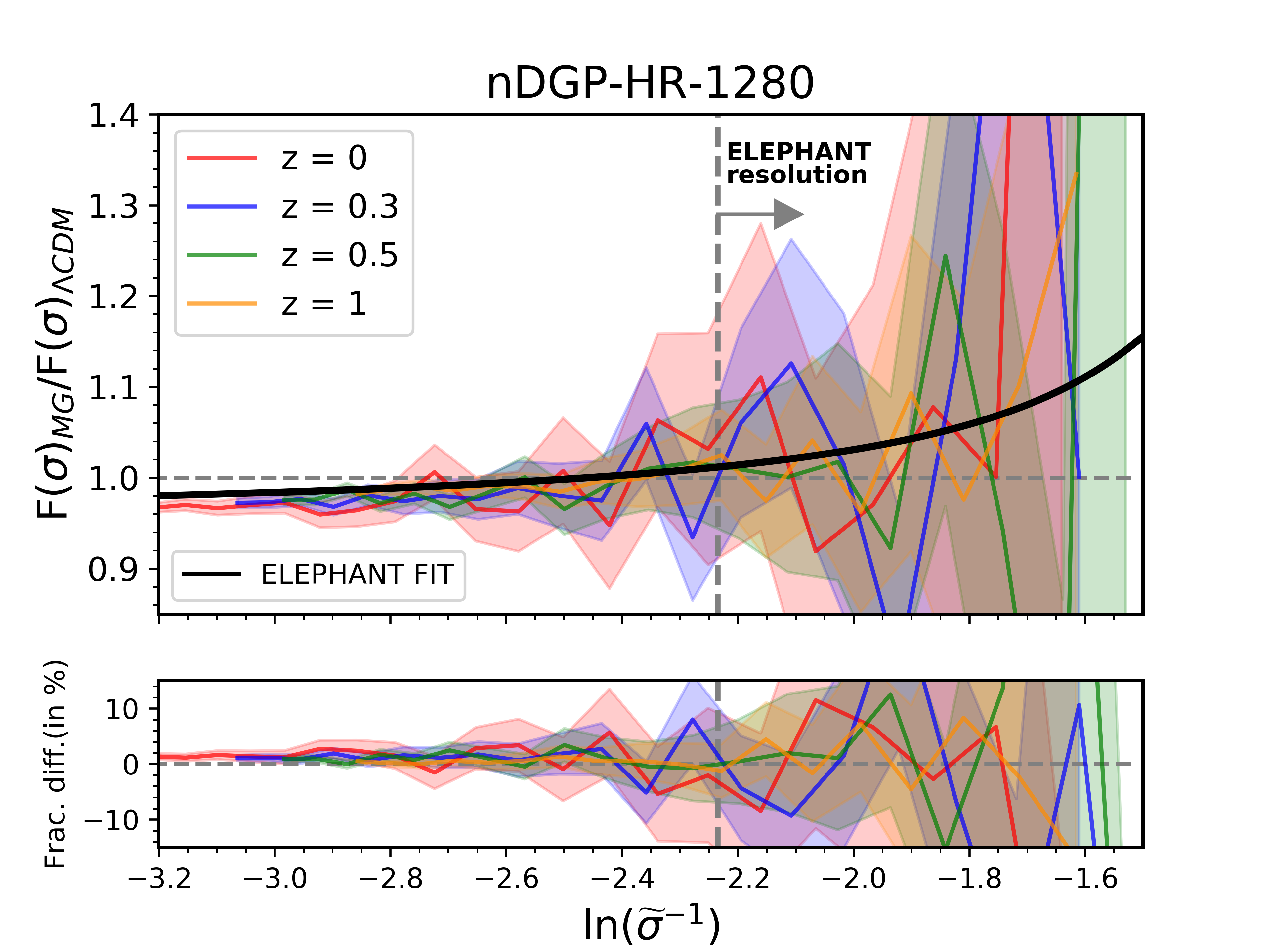}
 \caption{{\it Top panel:} The ratio, ${\fsig_{\text{nDGP(1)}}}/{\fsig_{\Lambda \text{CDM}}}$ obtained using \mgmil{} simulations for $z = 0, 0.3, 0.5$, and 1. The black curve represents the nDGP \elephant{} fit (Eq.\ \ref{eqn:fit_nDGP_equation}) with the best-fit parameters for nDGP(1) given in \cref{table:fit_nDGP_param}. {\it Bottom panel:} Percentage difference between the $\fsig$ values from the simulation and the corresponding values from the proposed fit. The grey vertical dashed lines in both panels illustrate the minimum ln($\sigma^{-1}$) accessible with the \elephant\ simulations.}
 \label{fig:f_n1_lcdm_mgmil}
\end{figure}

We start with \mgmil{}, which goes much deeper into the small-scale non-linear regime than our original runs. In \cref{fig:f_n1_lcdm_mgmil} we illustrate how our best fit of \cref{eqn:fit_nDGP_equation} performs in capturing the HMF deviations of nDGP(1) in this run. The bottom panel of this figure shows the percentage deviation in $\Delta_\text{MG}$ between the \mgmil\ simulations and our best-fit for this model, treated as a reference. The increased scatter in the \mgmil\ simulations
at the low-$\sigma(M)$ regime is expected, given nearly a 1000 times smaller volume of this run compared to \elephant{}. What is however outstanding is a remarkable agreement of our best-fit in the high-$\sigma$ range. Here, the differences are kept well below a few percent, even deep in the regime outside the original \elephant{}. The overall performance of our fit is very good. 

\begin{figure} 
 \includegraphics[width=\columnwidth]{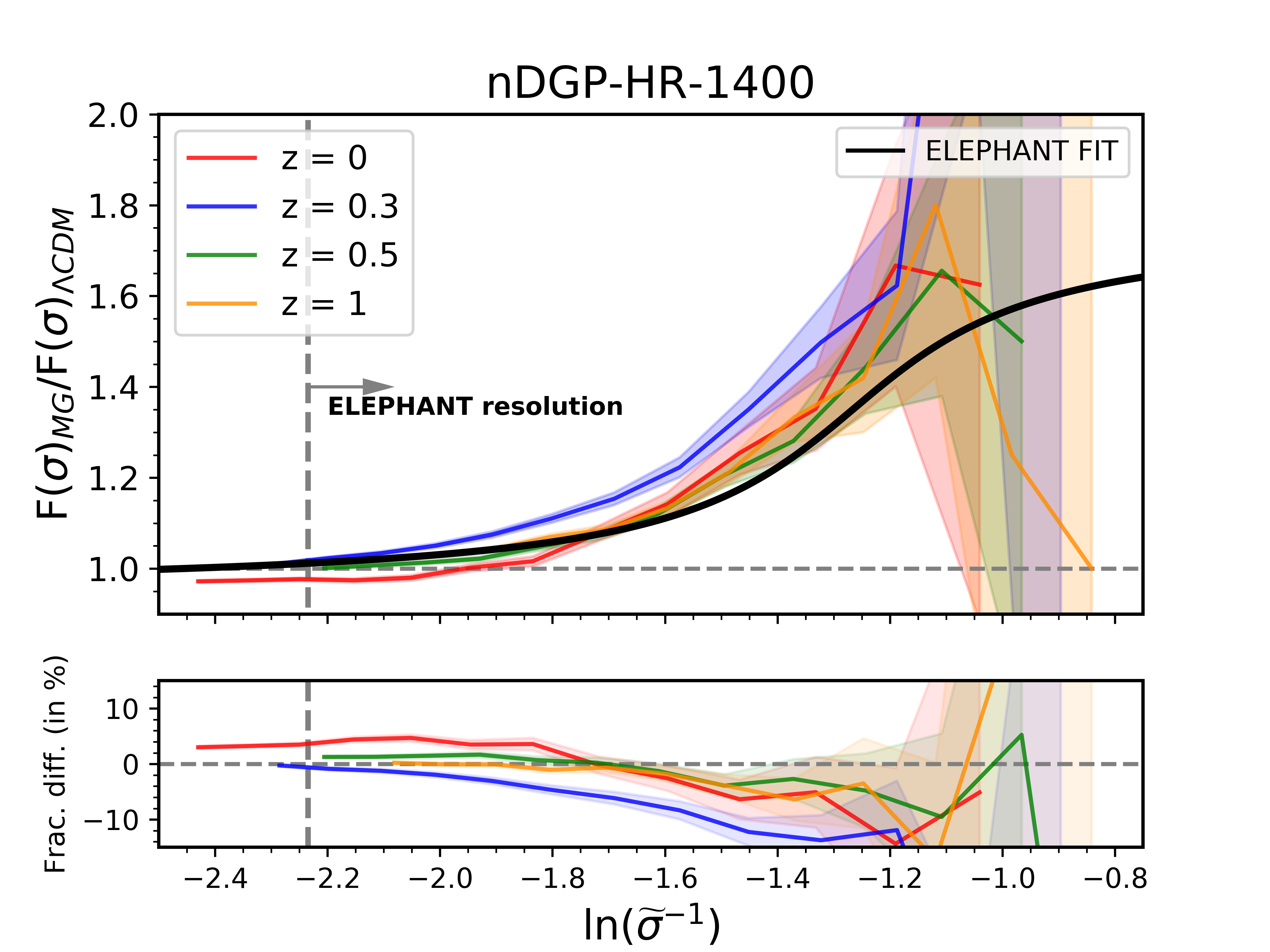}
 \caption{Analogously to \cref{fig:f_n1_lcdm_mgmil}, but for \dudeg{}.}
 \label{fig:f_n1_lcdm_dudeg}
\end{figure}

An analogous test for \dudeg{} is shown in \cref{fig:f_n1_lcdm_dudeg}. Here we observe that the agreement between our best-fit model prediction and the independent simulation data is worse than in the previous case. While the mismatch that we can see in the high-$\sigma$ range is still relatively small, usually staying within  5\%, the disagreement in the low variance regime is noticeably bigger. We note, however, that for this run also the overall degree of $\Delta_{\text{MG}}$ universality is substantially reduced. Still, the overall performance of our model is satisfactory as it stays within $\pm 10\%$ consistency with the data for the trusted ln$(\sigma^{-1})$ range, given the variance of \dudeg{} simulation runs.
%, this level of our model accuracy must be considered adequate. 

The results of the above tests reassure us that the universal nature of the $\Delta_\text{MG}(\sigma)$ we have found seems to be a real feature of the nDGP MG-class model. Moreover, it seems that the fitting formula we put forward for this gravity variant in \cref{eqn:fit_nDGP_equation} offers a very good and fully non-linear model of the MG HMF. However, the accuracy for the $f(R)$ family case (see Eq. \ref{eqn:fit_fR_equation}) would have to be further ascertained with high-resolution $f(R)$ runs. We leave this exercise for future work.

\section{Conclusions and discussion}
\label{sec:conclusions_discussions}

In this work, we have studied the dark matter halo mass function in modified gravity scenarios where structure formation differs from that in \lcdm. For that purpose, we employed the \elephant{} suite -- a set of $N$-body simulations, which cover GR and selected MG models, namely Hu-Sawicki $f(R)$ and nDGP. We focused on the intermediate to high-mass end of the halo  
distribution in the redshift range  $0 \leq z \leq 1$. In this regime, all the considered MG models display a redshift-dependent deviation in the HMF with respect to the \lcdm\ case, when analyzed as a function of halo mass.

We first verified that the MG HMFs as measured from simulations are not well matched by analytical models originating from the Press-Schechter framework \cite{PS74_MF}, such as the Sheth, Mo \& Tormen formula  \cite{S01_MF}, which describe the halo multiplicity function, $\fsig$. We attribute this failure of the analytical models to their ignorance of the complicated and inherently non-linear screening mechanism, which are a necessary ingredient in the cosmological MG scenarios, needed to satisfy observational constraints on gravity on both local scales and in high-energy conditions. 

%Comparison with the spherical collapse motivated models for the HMF, based on the Press-Schechter framework  \cite{PS74_MF}, showed that all of such models fail quite significantly to capture the MG HMF as predicted and measured in our N-body simulations. In this approach, one can predict the HMF amplitude across redshifts by modeling only the shape of the halo multiplicity function, $\fsig$, which is assumed to be universal across redshifts \cite{J01_MF}.

We note that the theoretical HMF models already show noticeable inaccuracies when contrasted with \lcdm{} $N$-body results. These inaccuracies are expected to be propagated, and most likely increased, when applied to MG models. To eliminate such leading-order discrepancies, instead of comparing the predictions of the absolute HMF amplitudes, we have focused on the relative ratios to the \lcdm-case, \ie\ $n_{\text{MG}}/n_{\Lambda\textrm{CDM}}$. The predictions for this ratio based on
the Sheth, Mo \& Tormen formula \cite{S01_MF} fails to capture the halo-mass dependent shape of the MG deviations as obtained from simulation results. 

For the $f(R)$-case, the SMT model under-predicts the HMF amplitude, while in contrast for the nDGP family we observe an over-prediction. This is again a clear manifestation of the shortcomings of these analytical models in capturing the extra MG physics, which is essentially related to the presence of intrinsically non-linear screening mechanisms operating at small scales in both models. %\SG{Screening of the fifth force is a necessary ingredient in the cosmological MG scenarios, needed to satisfy observational constraints on gravity on both local scales (\eg\ Solar System) and high-energy conditions (\eg\ gravitational wave sources).} For each model, this MG phenomenology introduces a new physical mass-scale separating, at a given epoch, the effectively screened halos from the unscreened ones.

In hierarchical structure formation scenarios, the abundance of collapsed objects is better characterized as a function of the rarity of the density peaks they originated from, rather than of a certain virial halo mass. The former is quantified by the density field variance $\sigma(M)$ at a given halo mass scale and redshift. Following this, we observed that $\fsig$ shows a much more universal character across redshifts for a given gravity model when expressed as a function of ln($\sigma^{-1}$). 
Furthermore, we have found that when we characterized the MG-induced effects as relative ratios of MG $\fsig$ to the \lcdm{} case, a new shape emerges, which is universal across redshift. While the deviations from \lcdm{} $\fsig$ for $f(R)$ models show a universal character already at their face-values, the nDGP case required an additional $\sigma(M)$ re-scaling. This extra step was needed to include the additional redshift-dependent magnitude of the fifth force in this class of models.

We have demonstrated that, once the MG HMF is expressed conveniently as a deviation from \lcdm{} case at a given $\sigma(M)$ scale, it exhibits a shape that is universal across redshifts. This is an important result, indicating that for models that employ such specific non-linear screening mechanisms, their effectiveness at the statistical level is well-captured by the filtering and expressing the density field in the natural units of its variance.

To better quantify and test this newly found universality, we invoked redshift-independent analytical fitting functions  to describe the $\Delta_{\text{MG}}\equiv F(\sigma)_{\text{MG}}/F(\sigma)_{\Lambda \text{CDM}}$ ratio. These fits were calibrated on the \elephant\ $N$-body simulations covering the redshift range from $z=0$ to $z=1$. For the $f(R)$ case,  we used a Gaussian-like form of the fitting function which captures a peak-like feature in $\Delta_{f(R)}$, %$\fsig_{f(R)}/\fsig_{\Lambda\text{CDM}}$,
the amplitude and position of which depends on the given $f(R)$ model's specifications. In the nDGP case, an arctan form proved to be a reasonably good fit for $\Delta_{\text{nDGP}}$,  %$\fsig_{\text{nDGP}}/\fsig_{\Lambda\text{CDM}}$, 
as it captures a monotonic increase at high-$\sigma$ (low-mass) end and suggests a limit of constant positive deviation at the high-mass range. Our best-fits turned out to provide quite good descriptions of the $\fsig$ for all tested MG variants, except for the $f4$ model data at $z=0$, which was a clear outlier. The fact that a single enclosed formula can provide a good fit for a given MG model at all redshifts reflects well that our $N$-body data supports the hypothesis of the redshift universality of $\Delta_{\text{MG}}$.

Using independent simulation runs with better resolution than the \elephant{} and a slightly different background cosmology, we were able to subsequently test our analytic approximations in the nDGP case. The level of agreement between our fits and these external data varied depending on the redshift and mass range, but overall it was satisfactory, with the departure of the fit well within 10\% of data-points in the trusted regime. This is a strong test indicating that the uncovered universal deviation of the HMF is a result of real physical phenomenology of the MG models in question, rather than a random chance effect unique to the particular \elephant{} suite. One could worry that a 10\% accuracy here is not an impressive precision, given that $1\%$ statistical precision will be demanded by the forthcoming Big-Data cosmological surveys. However, on the simulation side, it has been shown that the agreement between present day different halo finders is at most $10\%$ in \lcdm{} \citep{Halos_MAD,desi_nbody_comparison}. Thus our
accuracy reported here is already approaching the current numerical limit. On the other hand, the magnitude of deviations from GR as fostered by our MG models typically reaches a factor a few$\times 10\%$ at $\ln(\sigma^{-1})$ values where our accuracy limit is set.

The resolution of the \elephant{} simulations allowed us to robustly probe only intermediate and large mass halos. In this limited mass-$\sigma$ regime, the abundance of structures in MG increases w.r.t.\ \lcdm{}, as small mass halos accrete and merge faster to form larger structures. However, owing to the conservation of mass in the universe, we can expect that there should be a simultaneous decrease in the number of small-mass halos in the MG models when compared to \lcdm{}. This was found to be indeed the case for some of MG variants  \citep[\eg][]{borderline_f6_hmf,HMF_REBEL,hmf_ndgp_cluster}. A similar effect is also hinted in our results of the high resolution \mgmil{} runs at small ln($\sigma^{-1}) < -2.75$. % in \cref{fig:f_n1_lcdm_mgmil}. 
Thus, a natural extension of our study and an important further test of the $\Delta_{\text{MG}}$ universality will be to probe a smaller mass (and larger $\sigma$) regime. Such a study will require a completely new set of high-resolution $N$-body simulations, and we plan it as a future project.

The HMF and its time evolution is one of the most important and prominent predictions of the theory of gravitational instability and formation of the large-scale structures  \cite{peebles_1980}. In the \lcdm{} framework and within its GR-paradigm, there is strong evidence for the universality of the HMF with redshift, when the HMF is expressed in the units of the dimensionless cosmic density field variance. Once we admit a model with an extra fifth force acting at intergalactic scales, such as the MG models studied here, the universality of the HMF could no longer be taken for granted. We have shown that once the density field is re-scaled, both of our MG models exhibit an approximately universal $\fsig$, similar to the trend seen in the case of \lcdm{}. Moreover, its ratio w.r.t.\ the \lcdm{} case can now be modeled by a single enclosed formula, with a good fit for each of the specific $f(R)$ and nDGP model variants. This opens an avenue for building accurate, yet relatively simple, effective models of the HMF in MG scenarios. Such models can then be implemented in studies on galaxy-halo connection, galaxy bias, and non-linear clustering \citep[\eg][]{Schmidt_Oyaizu_2008_3,hm_galileon,hm_nonlocal_gravity,sphericalcollapse_braneworld,Lombriser_2013,fR_nonlinear_structure,halomod_fR,hm_chameleon}.
This will be of paramount importance for %forecasting and yielding 
robust predictions on cosmological observables and their covariance. %using methods such as Monte-Carlo-Markov-Chain simulations. 
Such calculations for many observables of interest have been so far severely limited, mainly because $N$-body simulations, with a full non-linear implementation of screening mechanisms, are prohibitively expensive for the case of most non-trivial MG scenarios. 

From this standpoint, the results of our analysis are the first step towards building a versatile, accurate, and numerically cheap model of non-linear matter and galaxy clustering for MG cosmologies. 

%\SG{This could} \MB{provide a ?great? alternative} %serve complimentary to the development of 
%\SG{to other fast methods, like emulators} %\citep{emulator_main,emulator_main2}}
%\MB{, aimed at testing} %for predicting and designing future experiments to test 
%\SG{the signatures of modifications to gravity in cosmological observables} \MB{from today's and future experiments.}% \SG{\citep {emulator_pk_fR,emulator_FORGE,emulator_pk_arbitrary,emulator_pk_beyond_lcdm}. Emulators are an efficient tool to make estimates and study the impact of cosmological parameters on non-linear observables \citep{emulator_coyote,emulator_dark,emulator_hmf_aemulus,emulator_hmf_mira,cosmic_emulation}. 
%\MB{Semi-analytical models for beyond-GR scenarios, combined with the predictions from emulators, along with $N$-body simulations, would enable us to achieve high precision,} \MB{necessary to reliably} %, which would be needed to
%\SG{forecast constraints and analyze results from ongoing and future surveys.} 

\begin{acknowledgments}
The authors would like to thank Hans A. Winther for kindly providing us with \textsc{mgcamb} version with specific forms of $\mu(a,k)$ and $\gamma(a,k)$ functions implementing our $f(R)$ and nDGP models. We also thank the anonymous referee whose comments help improve this manuscript. 
This work is supported via the research project ``VErTIGO'' funded by the National Science Center, Poland, under agreement no. 2018/30/E/ST9/00698. We also acknowledge the support from the Polish National Science Center within research projects no. 2018/31/G/ST9/03388, 2020/39/B/ST9/03494 (WAH \& MB), 2020/38/E/ST9/00395 (MB), and the Polish Ministry of Science and Higher Education (MNiSW) through grant DIR/WK/2018/12. This project also benefited from numerical computations performed at the Interdisciplinary Centre for Mathematical and Computational Modelling (ICM), University of Warsaw under grants no. GA67-17 and  GB79-7.
\end{acknowledgments}

\appendix*
\section{Re-scaling the matter variance in nDGP gravity}
\label{appendix:rescale_nDGP}

\begin{figure*}
\includegraphics[width=\columnwidth]{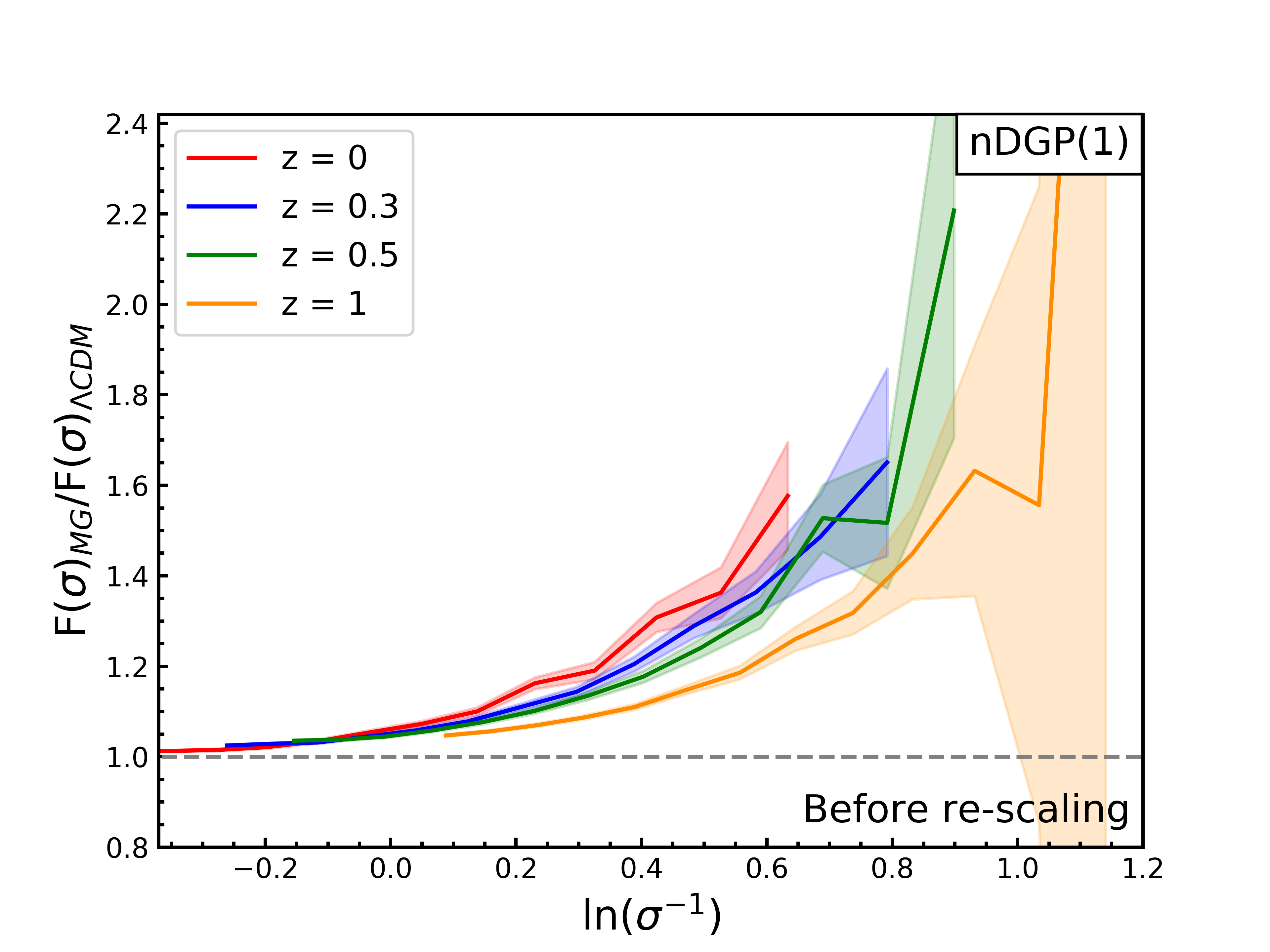}
\includegraphics[width=\columnwidth]{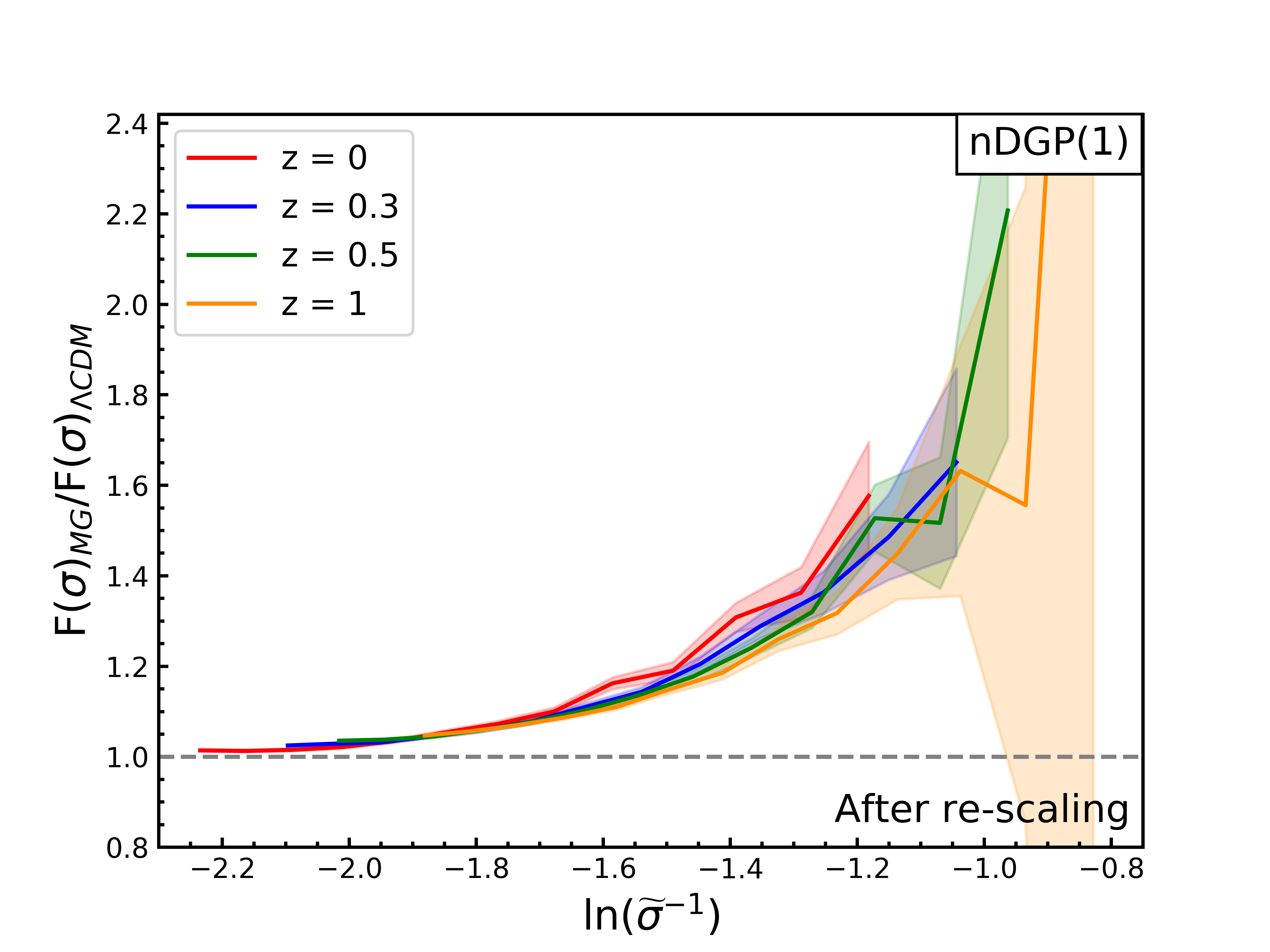}
 \caption{\textit{Left:} The ratio $\frac{\fsig_{\text{nDGP(1)}}}{\fsig_{\Lambda \text{CDM}}}$ as a function of ln($\sigma^{-1}$).
 The trend becomes less universal as the scales increase and clearly depends on redshift at larger values of ln($\sigma^{-1}$).
 \textit{Right:} The same ratio as to the left, but after re-scaling the matter variance, by including
 the force enhancement term, $\Xi(z)$, in  $\ln(\widetilde{\sigma}^{-1})$, where $\widetilde{\sigma}(z) \equiv \sigma/\Xi(z)$. The resultant plot shows a comparatively more universal trend across epochs. } 
\label{fig:n1_scaling_noscaling}
\end{figure*}

Force enhancement due to the scalar field gradient in the case of nDGP gravity is given by:
\begin{equation}
\Xi(z) \equiv \frac{F_{5th}}{F_N} = \left(\frac{\textrm{d}\phi}{\textrm{d}r}\right)\bigg/\left(\frac{\textrm{d}\Psi_N}{\textrm{d}r}\right)\,,    
\end{equation}
where $\phi$ is the scalar degree of freedom associated with the fifth force, and $\Psi_{N}$ is the standard (\ie~ Newtonian) gravitational potential. 
Considering the Vainshtein screening for a spherically symmetric body with a Lagrangian radius, $R_L$, $\Xi(z)$ is given by     \cite{sphericalcollapse_braneworld, dynamical_mass_mg, nonlinear_interactions_nDGP}:
\begin{equation}
\Xi(z) =  \frac{2}{3\beta}\left(\frac{R_L}{r_{v}(z)}\right)^{3}\left(\sqrt{1+\left(\frac{R_L}{r_{v}(z)}\right)^{-3}}-1 \right),
\label{eqn:enh_nDGP}
\end{equation}
where the Vainshtein radius $r_{v}(z)$ is :
\begin{equation}
\label{eqn:vainshtein_radius}
    r_{v}(z) = \left( \frac{16r_{c}^{2}Gm(r)}{9\beta(z)^{2}} \right)^{1/3}
\end{equation}
and
\begin{equation}
    \label{eqn:beta_ndgp}
    \beta(z) = 1+2H(z)r_{c}\left(1+\frac{\dot{H}}{3H(z)^{2}}\right).
\end{equation}

As the cross-over scale $r_{c}$ increases, $r_{v}(z)$ becomes larger and $\Xi(z)$ in \cref{eqn:enh_nDGP} goes to zero, thereby screening the fifth force and recovering GR. Since the Vainshtein radius depends on redshift, this makes the force enhancement factor in nDGP an
intrinsically time-dependent function.

We used the formula for $\Xi(z)$ to remove this first-order intrinsic time-dependent enhancement in nDGP w.r.t.\ the GR-case, by considering re-scaled matter variance $\widetilde{\sigma}(z) \equiv \sigma(z)/\Xi(z)$.

In the left-hand plot of \cref{fig:n1_scaling_noscaling} we have plotted the quantity $\frac{\fsig_{\text{nDGP(1)}}}{\fsig_{\Lambda \text{CDM}}}$ without re-scaling, and we see explicit dependence of this ratio on redshift. After re-scaling of matter variance in the right plot, we can acknowledge the resultant universal ratio of the nDGP(1) HMF w.r.t.\ \lcdm{} across redshifts. A similar trend is seen in the case of nDGP(5) and we show the resultant re-scaled plots in the main text.

\bibliography{universal_hmf_mg}

\end{document}